\documentclass[graybox]{svmult}

\usepackage{type1cm}        
\usepackage{makeidx}         
\usepackage{graphicx}        
\usepackage{multicol}        
\usepackage[bottom]{footmisc}

\usepackage{newtxtext}       %
\usepackage[varvw]{newtxmath}       

\usepackage{hyperref}
\hypersetup{
    colorlinks=true,
    citecolor=blue,
    linkcolor=blue,
    filecolor=magenta,      
    urlcolor=blue,
    }

\usepackage[most]{tcolorbox}

\makeindex             

\usepackage{natbib}    

\def\civ     {\ensuremath{\text{C\,\textsc{iv}}}}

\def\mgii     {\ensuremath{\text{Mg\,\textsc{ii}}}}

\def\cii     {\ensuremath{\text{[C\,\textsc{ii]}}}}

\def\halpha {\ensuremath{\text{H}\alpha}}

\def\nii     {\ensuremath{\text{[N\,\textsc{ii]}}}}

\def\neiii     {\ensuremath{\text{[Ne\,\textsc{iii]}}}}

\def\oii     {\ensuremath{\text{[O\,\textsc{ii]}}}}

\def\oiii     {\ensuremath{\text{[O\,\textsc{iii]}}}}

\begin{document}

\title*{Observations of Early Black Holes Before and After JWST}
\author{Eduardo Ba\~nados}
\institute{Max-Planck-Institut f\"ur Astronomie, K\"onigstuhl 17, 69117 Heidelberg, Germany, \email{banados@mpia.de}}

\maketitle

\abstract{
These notes are from three lectures given at the 54th Saas-Fee Advanced Course of the Swiss Society of Astrophysics and Astronomy in January 2025. This chapter reviews the dramatic evolution in our understanding of supermassive black holes in the first billion years, from ground-based discoveries to recent space-based infrared observations with JWST. Section 1 introduces AGN and quasars to contextualise observations at the highest redshifts. Section 2 reviews the pre-JWST understanding of early quasars, including personal accounts of how key discoveries were made. Section 3 examines how JWST is transforming the field, from black hole mass measurements and host galaxy characterisation to large-scale environmental studies, and identifies emerging directions. }

\section*{Preface} 
 My goal here is not to create a comprehensive textbook but rather to review the main ideas, point you to additional literature, and share some personal anecdotes.    A brief disclaimer: my lectures focus on observational aspects and inevitably reflect my own biases.

While preparing these lectures, I thought about what I would have liked to know as a graduate student---connections that took me too long to piece together. The result is a set of material that, although certainly incomplete, I believe provides essential context for  understanding current high-redshift observations.

These lectures reflected the state-of-the-art as of January 2025, when the 54th 
Saas-Fee Advanced Course took place. Given the rapid pace of this 
field, I could not resist updating beyond what I originally presented. I have 
tried to remain faithful to the topics covered in the school, but Section~\ref{banados-sec:jwst} in 
particular includes results that appeared between the lectures and the completion 
of these notes. Even so, a further update will 
likely be needed before long. 

\section{Introduction to Early Black Holes and Their Observational Signatures}
\label{banados-sec:intro}

To appreciate both the remarkable advances and persistent limitations in studying supermassive black holes at cosmic dawn, we must begin closer to home with the physical phenomena that define Active Galactic Nuclei (AGN) and quasars. 
In this section, I cover the fundamentals: historical nomenclature, key definitions, observational signatures, and methods for measuring black hole masses.

\subsection{Active Galactic Nuclei and Quasars: Fundamental Concepts}

I started my first lecture by asking, ``Without consulting \textsc{ChatGPT}  \citep{chatgpt2024}, what are AGN and Quasars?", fearing that nobody would raise their hands. 
 To my pleasure, the students were extremely enthusiastic and already had excellent intuitions about these objects.  In the era of Artificial Intelligence (AI) Large Language Models (LLM), I was curious about their answer to the question. \textsc{ChatGPT} provided the following response:

\begin{quotation}
Active Galactic Nuclei (AGN) and quasars are among the most luminous and energetic phenomena in the universe, powered by supermassive black holes (SMBHs) at the centers of galaxies. 
\end{quotation}

I then asked, ``What is the difference between AGN and quasars?'' And this was its answer: 

\begin{quotation}
The difference between AGN (Active Galactic Nuclei) and quasars lies in their luminosity and context within the broader category of active galaxies. Here's a detailed comparison: 
\end{quotation}

\begin{figure}[t]
\includegraphics[width=\textwidth]{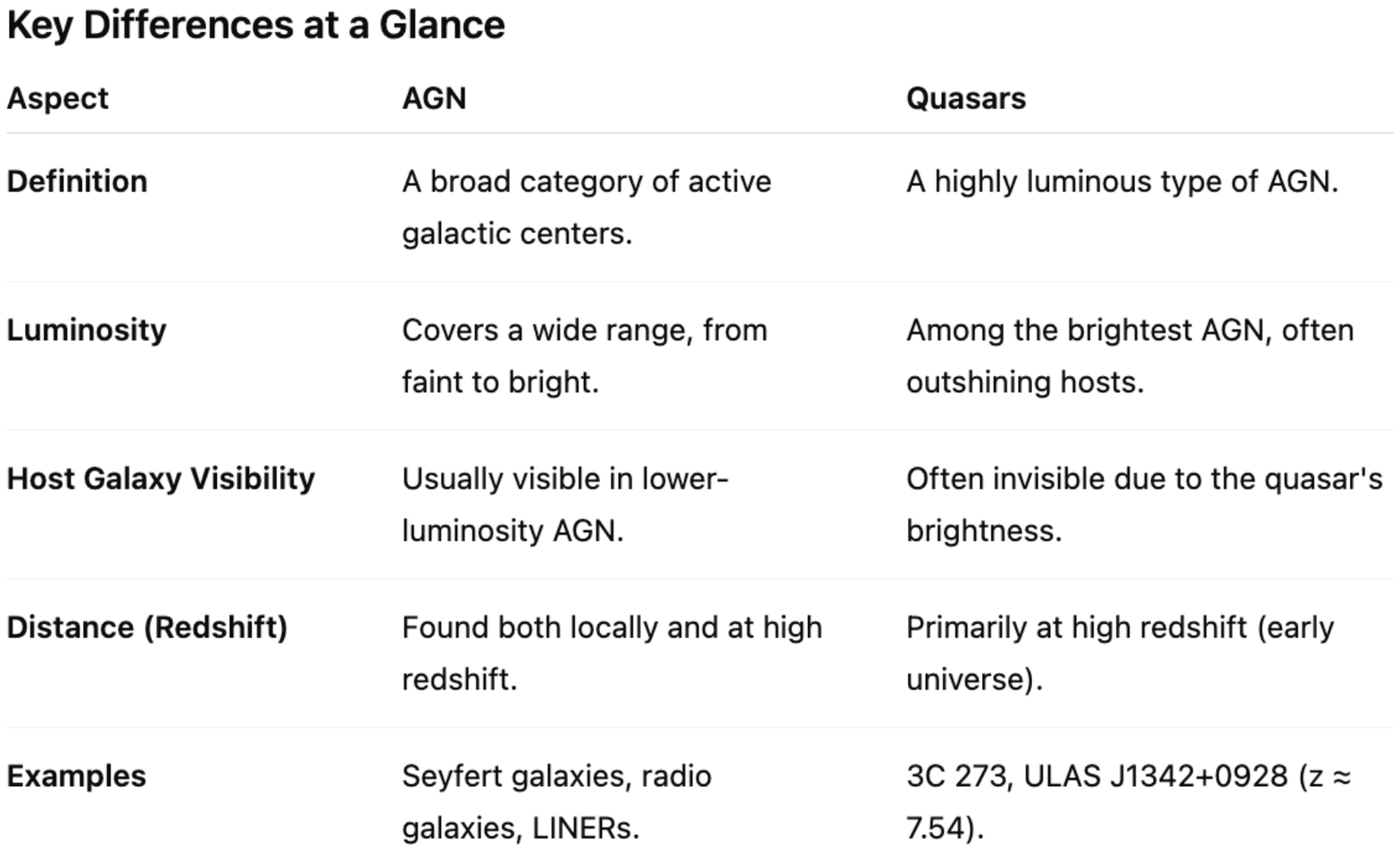}
\caption{\textsc{ChatGPT}'s comparison table generated in response to the prompt ``What is the difference between AGN and quasars?'' The AI correctly identifies luminosity as the primary distinction. Note that the ``Distance'' entry was accurate before JWST observations but has since been superseded by recent discoveries.}
\label{fig:chatGPTagn}     
\end{figure}

The \textsc{ChatGPT} table comparison for AGN and Quasars is shown in Fig.~\ref{fig:chatGPTagn}. This was a better outcome than I had imagined; it is something I could have said myself and reflected very well the students' answers. I was particularly surprised to see \textsc{ChatGPT} cite two particular quasars, one  that will be described in more detail in this chapter: 3C\,273 \citep{schmidt1963}, the first identified quasar, and the ``Pisco'' ULAS~J1342+0928 quasar, which I discovered \citep{banados2018a}. 

The key insight, correctly identified by \textsc{ChatGPT}, is that the main distinction is luminosity: quasars are simply the most luminous type of AGN (see Fig.~\ref{fig:luminosity}). 
However, this exercise also revealed important gaps in AI understanding and knowledge. Most critically, AGN represent \textit{a phase of a galaxy }—moments when we observe the central supermassive black hole actively growing. This temporal aspect is fundamental but often overlooked in static definitions. 

Additionally, \textsc{ChatGPT}'s ``Distance'' comparison in Fig.~\ref{fig:chatGPTagn} was accurate before JWST, but as we will learn, recent observations have uncovered AGN at higher redshifts than the most distant known quasars, blurring this traditional distinction. 

Rather than rely on AI summaries, let's build this understanding from first principles, starting with the historical context that led to our current paradigm.

\begin{backgroundinformation}{Practical AGN and Quasar Classification}
Significant confusion exists in the literature regarding AGN and quasar nomenclature.  Since AGN activity represents a transient phase, the same object can change categories even on human timescales, while inconsistent definitions in the literature compound the confusion. 

For practical purposes, I use Fig.~\ref{fig:luminosity} as a guide to distinguish between galaxies, AGN, and  (faint and bright) quasars. 
The brightest UV sources ($M_{1450} < -23$) are uncontroversially classified as quasars. At intermediate luminosities ($-23 < M_{1450} < -21$), objects are often called ``faint quasars'' by some authors and ``AGN'' by others. At the lowest luminosities ($M_{1450} > -21$), distinguishing AGN from luminous star-forming galaxies requires additional information beyond simple luminosity cuts (JWST is expanding the classification toolkit; see e.g., Section~\ref{subsec:line_diagnostics}).
\begin{figure}[t]
\includegraphics[width=\textwidth]{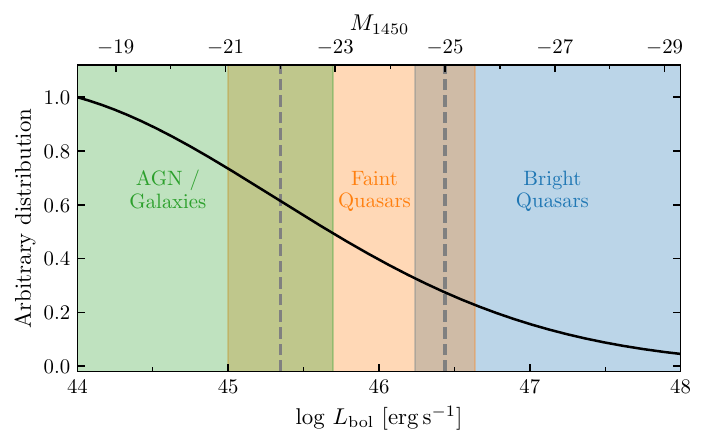}
\caption{Practical luminosity boundaries for classifying AGN and quasars. The coloured regions indicate approximate divisions based on the UV absolute magnitude $M_{1450}$ (top axis) and bolometric luminosity (bottom axis), with the conversion between the two axes following \cite{runnoe2012}. The black curve illustrates the typical luminosity distribution. While these boundaries are somewhat arbitrary, they provide useful guidance for the literature, where classification schemes often vary between studies.}
\label{fig:luminosity}    
\end{figure}

\end{backgroundinformation}

\subsection{AGN and Quasars: A Historical Perspective}
\label{banados-subsec:agndef}

Nowadays, we take for granted that AGN and quasars are powered by supermassive black holes — but this understanding emerged only after decades of debate and discovery. The journey from mysterious ``quasi-stellar objects'' to our modern picture of accretion-powered black holes provides essential context for interpreting observations in the early Universe. 
Many books have been written about the history of ``black holes'' \citep[a personal favourite: ][]{bartusiak2016}---here I will briefly summarise the ``historical papers'' discussed in the lecture. 

In hindsight, astronomers had been observing AGN for decades without understanding their true nature. As early as 1943, Carl Seyfert identified a class of spiral galaxies with unusually bright nuclei and exceptionally broad emission lines---what we now recognise as Seyfert galaxies, a subset of the AGN family \citep{seyfert1943}. Similarly, in 1954, Baade and Minkowski demonstrated that Cygnus A, one of the brightest radio sources in the sky, was an extragalactic object showing the first clear evidence of extragalactic radio jets \citep{baade1954}. Yet the physical mechanism powering these phenomena remained a complete mystery.

A breakthrough came in 1963 when Maarten Schmidt was taking spectra of ``radio stars'' from the third Cambridge radio catalogue. The spectrum of the radio source 3C~273 did not match a stellar spectrum; then he realised that its puzzling emission lines were familiar hydrogen Balmer features, redshifted by $z=0.16$ \citep{schmidt1963} — placing this star-like object at cosmological distances with an implied luminosity more than 100 times higher than the entire Milky Way. 

\begin{figure}[h]
\includegraphics[width=0.95\textwidth]{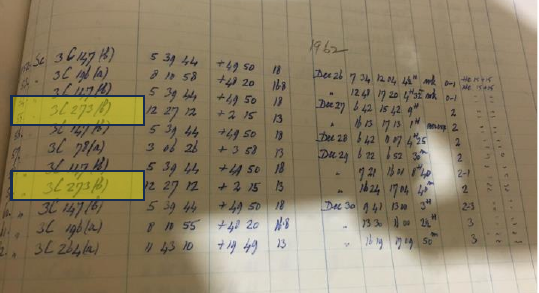}
\caption{Maarten Schmidt's original observing log that resulted in the now famous discovery of the first quasar 3C~273 \citep{schmidt1963}. This photo was taken in 2016 at the Carnegie Observatories' basement vault.}
\label{fig:3c273log}     
\end{figure}

During my time as a postdoc at Carnegie Observatories, I had the privilege of finding Schmidt's original observing logs for 3C~273, which felt like uncovering a hidden treasure (Fig.~\ref{fig:3c273log}). I was also fortunate to meet Maarten Schmidt himself after giving a colloquium at Caltech in 2017, a tangible and memorable connection to this pivotal moment when quasars were first recognised as a new class of cosmic phenomena.

Schmidt's discovery unleashed a flood of similar findings. By 1964, the term ``quasar'' (quasi-stellar radio source) had been coined \citep{greenstein1964}. However, it soon became clear that many such objects lacked strong radio emission, leading to the broader classification of quasi-stellar objects, or QSOs \citep{sandage1965}. Nowadays, both ``quasar'' and ``QSO'' are used interchangeably for the same phenomena. 

The theoretical framework for understanding these luminous engines emerged gradually. \cite{salpeter1964} first proposed that accretion onto massive compact objects could provide the necessary power. In a landmark paper, \cite{lynden-bell1969} explicitly connected quasars to supermassive black holes and argued that galaxy centres should contain dormant remnants of past quasar phases. The foundation for accretion disk physics was laid by \cite{shakura1973}, whose model remains central to our understanding of accretion theory.

The fundamental questions that puzzled Schmidt and his contemporaries---how such compact objects can sustain extreme luminosities, what physical processes govern their activity, and how they evolve over cosmic time---remain at the heart of our investigations into supermassive black holes in the early Universe.

\subsection{Key Observational Properties: Why Do We Think an SMBH Is at the Centre of an AGN?}
\label{sec:key_properties}

The identification of supermassive black holes as the central engines powering AGN rests on compelling observational properties that, when considered collectively, point to no other plausible explanation.

\begin{figure}[h]
\includegraphics[width=\textwidth]{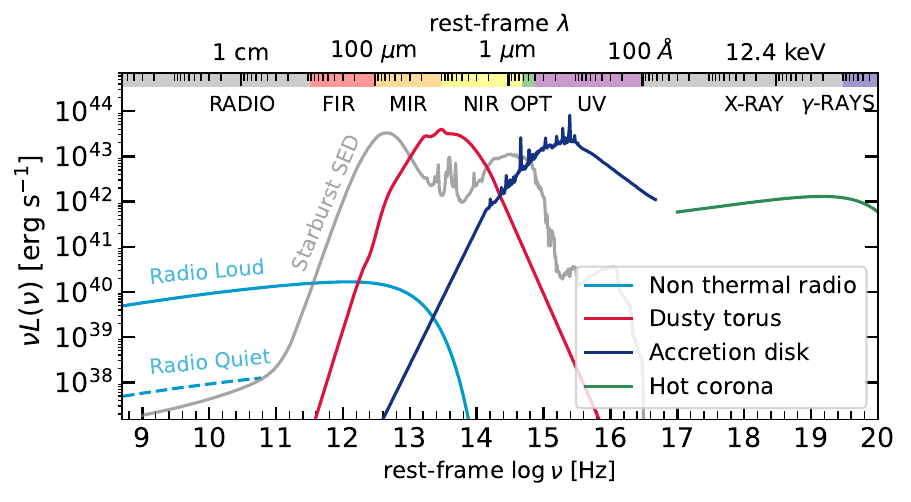}
\caption{Schematic spectral energy distribution of an AGN (coloured curves) compared to a starburst galaxy (grey curve). Different physical processes around the supermassive black hole contribute distinct spectral components that cannot be reproduced by stellar populations alone. Figure created with \textsc{AGNFITTER-RX} \citep{martinez-ramirez2024}, courtesy of L.~Mart\'inez-Ram\'irez.}
\label{fig:agnsed}    
\end{figure}

\begin{figure}[h]
\sidecaption
\includegraphics[width=0.62\textwidth]{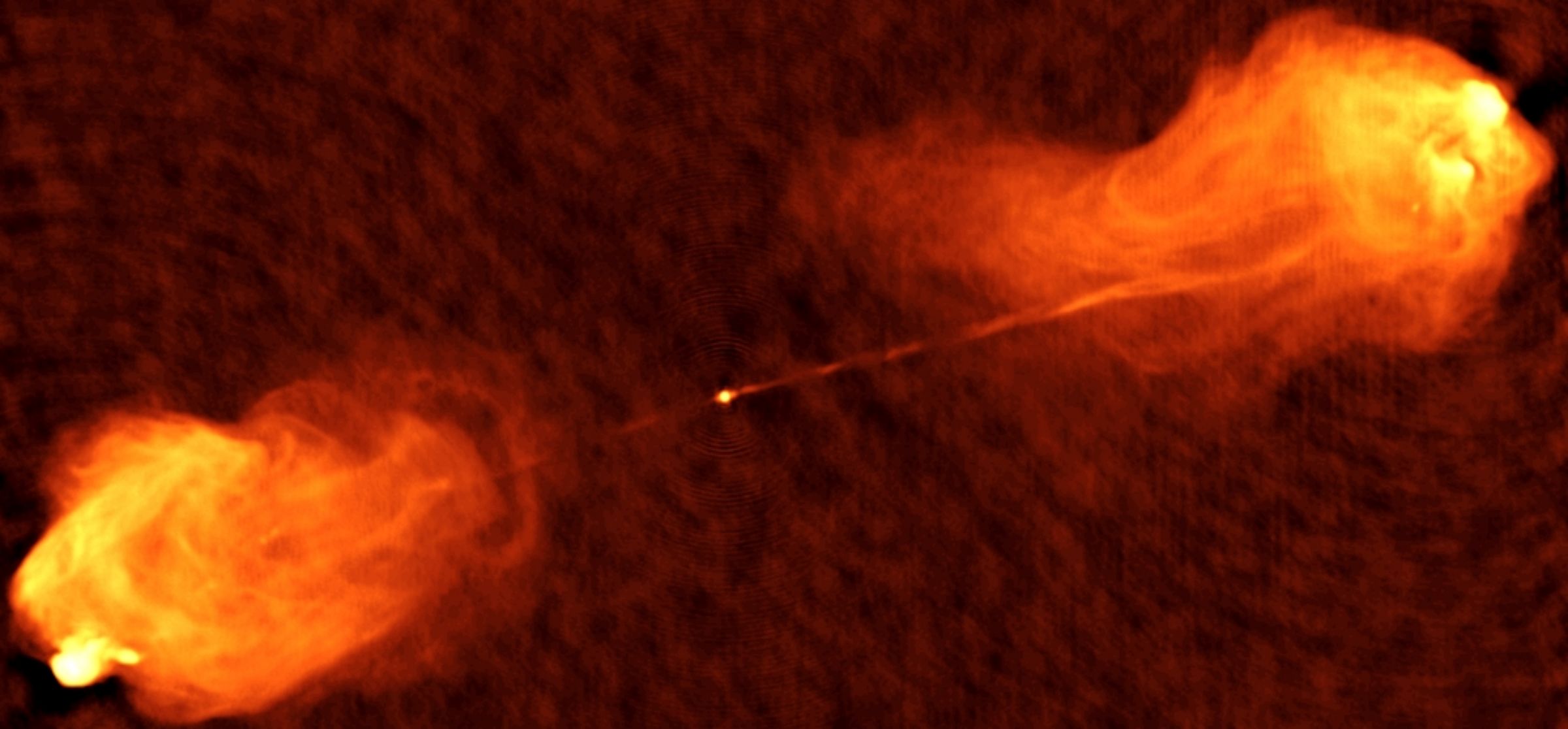}
\caption{Multi-frequency radio observations of Cygnus A \citep{perley1984}, showing the classic double-lobed structure with jets extending over 100\,kpc from the central black hole. This is one of the brightest radio sources in the sky and the first identified extragalactic jet \citep{baade1954}. Cygnus A has recently been studied also by JWST \citep{ogle2025}. 
Image credits: \href{https://www.nrao.edu/archives/items/show/33386}{NRAO/AUI Archives}. }
\label{fig:cygnus}    
\end{figure}

\begin{figure}[h]
\includegraphics[]{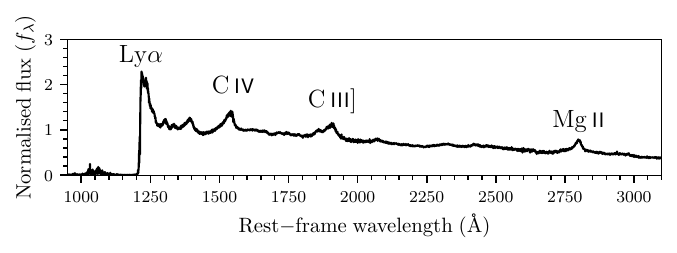}
\caption{Average rest-frame UV spectrum of the brightest $z\sim 6$ quasars \citep{dodorico2023}. The spectrum shows the characteristic broad emission lines (labelled) with Doppler widths reaching $\sim$10\,000\,km\,s$^{-1}$, indicating gas in rapid orbital motion around supermassive black holes. The underlying continuum shows power-law emission from the accretion disk.}
\label{fig:xqr30}    
\end{figure}

\begin{figure}[h]
\sidecaption
\includegraphics[width=0.63\textwidth]{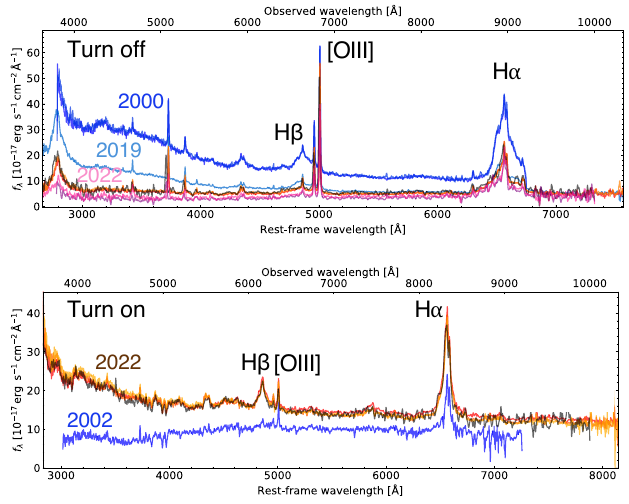}
\caption{Examples of changing-look quasars showing dramatic spectral evolution on human timescales. \textbf{Top}: A quasar ``turning off'' between 2020 and 2022, with broad H$\beta$ and H$\alpha$ emission disappearing. \textbf{Bottom}: A quasar ``turning on'' between 2002 and 2022, with broad emission lines appearing. These transitions occur over months to years, demonstrating that AGN are events or phases of a galaxy.  Adapted from \cite{zeltyn2024}.}
\label{fig:CLQSO}    
\end{figure}

\begin{enumerate}
\item{\textbf{Extreme Luminosities from Compact Regions}.
AGN exhibit extraordinarily large luminosities from highly compact regions. The most luminous quasars can reach bolometric luminosities of $\sim$$10^{14}\,L_\odot$, yet this enormous energy output originates from regions smaller than a few parsecs. 
This combination of extreme luminosity and compactness immediately rules out stellar processes as the primary energy source—only gravitational potential energy from matter falling into a supermassive black hole can account for such extreme energy densities.
}
\item{\textbf{Distinctive Spectral Energy Distributions}. 
The spectral energy distributions (SED) of AGN are fundamentally different from stellar populations, exhibiting a complex multi-component structure across the electromagnetic spectrum. A scheme of an AGN SED is displayed in Fig.~\ref{fig:agnsed}. Key components include the distinctive ``big blue bump'' in the UV/optical from thermal accretion disk emission, hard X-ray emission from hot coronae, mid-infrared emission from heated dust in a circumnuclear torus, and, in some sources, radio spectra from synchrotron jets. This multi-component SED, with different physical processes dominating different wavelength ranges, cannot be reproduced by stellar populations alone and requires the extreme conditions present near supermassive black holes.
}
\item{\textbf{Relativistic Jets.}
Many AGN exhibit highly collimated jets that move at relativistic speeds, maintaining their collimation over scales ranging from parsecs to megaparsecs (see Fig.~\ref{fig:cygnus}). VLBI observations reveal apparent superluminal motion\footnote{See movies showing superluminal motion \href{https://personal.denison.edu/\%7Ehomand/superluminal/}{in this link}} with Lorentz factors $\Gamma > 10$, providing direct evidence for bulk relativistic motion (e.g., \citealt{lopez-corredoira2012,lister2018}). The physics of jet launching requires the extraction of rotational energy from a spinning black hole through the  Blandford-Znajek mechanism \citep{blandford1977}---no other astrophysical object can sustain such powerful, stable jets.}
\item{\textbf{Extreme Doppler-Broadened Emission Lines.}
AGN spectra show emission lines with extreme Doppler broadening reaching velocities up to $\sim$$10\,000$\,km\,s$^{-1}$ (Fig.~\ref{fig:xqr30}). These broad lines originate from gas in rapid orbital motion around the central mass concentration within $\sim$1 parsec of the central engine. The combination of high velocities and small distances directly constrains central masses of 10$^{6-10}$\,$M_\odot$.}
\item{\textbf{Rapid Variability Across All Wavelengths.}
AGN exhibit dramatic variability across all wavelengths on timescales from minutes to decades \citep[e.g.,][]{fernandes2020,kozlowski2019}. 
The shortest variability timescales directly constrain emission region sizes to light-hours, consistent with the innermost stable circular orbit around a supermassive black hole. ``Changing-look'' quasars exemplify this dramatic behaviour, transitioning between different AGN types over months to years as their broad emission lines appear or disappear \citep[][see Fig.~\ref{fig:CLQSO}]{ricci2023}. }
\end{enumerate}

While each property individually challenges alternative explanations, their combination makes the case for supermassive black holes compelling. Any alternative model must simultaneously account for all five observational signatures described above. The supermassive black hole model, combined with accretion disk physics and general relativistic effects, naturally accounts for all observations within a single, coherent theoretical framework. This observational foundation provides crucial context for interpreting JWST's revolutionary observations as we probe the earliest epochs of supermassive black hole growth and identify the first generation of quasars in the infant universe.

\subsection{The AGN Unification Paradigm}
\label{subsec:agn_unification}

The historical zoo of AGN classifications—Seyfert galaxies, quasars, blazars, radio galaxies—were once regarded as distinct phenomena. However, the modern unification paradigm suggests that many of these differences arise primarily from orientation effects and the varying presence of obscuring material rather than intrinsic physical distinctions \citep{Antonucci1993,Urry1995}. 

\begin{figure}[h]
\includegraphics[width=\textwidth]{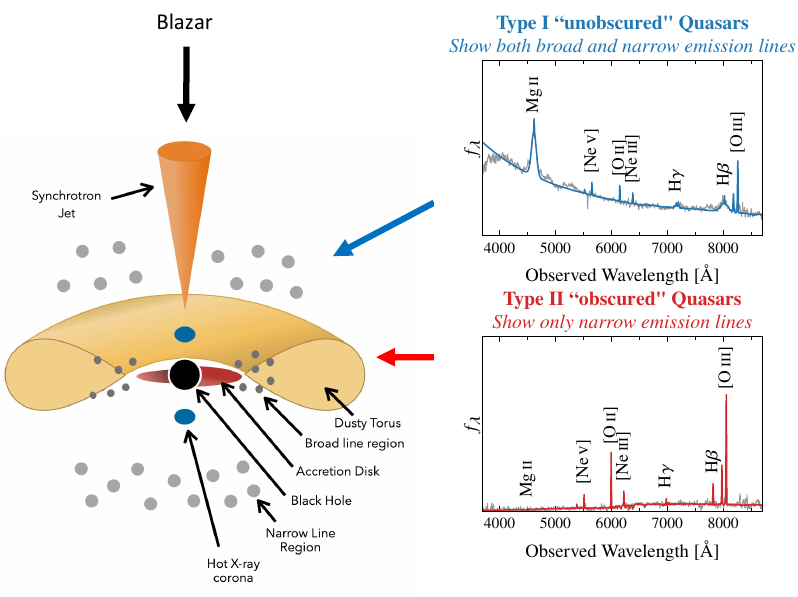}
\caption{Schematic representation of the AGN unification model illustrating the orientation-dependent appearance of AGN. \textbf{Left}: Cross-sectional view showing the key structural components. 
The diagram exhibits vertical symmetry, except for the relativistic radio jet, which indicates that only a fraction of AGN possess a powerful radio jet. Arrows indicate different viewing angles corresponding to various AGN classifications. 
The smooth dusty torus geometry displayed here is a simplification; observational evidence increasingly supports clumpy, cloud-based structures \citep[see e.g.,][]{elitzur2007,honig2019}.
\textbf{Right}: Representative optical spectra demonstrating the observational consequences of the unification paradigm. Type I unobscured quasars (blue arrows, moderate inclination) exhibit both broad (FWHM $>$1000\,km\,s$^{-1}$) and narrow emission lines, while Type II obscured quasars (red arrows, edge-on viewing) show only narrow emission lines due to obscuration of the broad line region by the dusty torus. Schematic diagram adapted from \cite{thorne_2022_6381013}; spectra from \cite{hviding2024}.  }
\label{fig:unification}     
\end{figure}

Figure~\ref{fig:unification} shows a schematic diagram of the key structural components of the AGN unification model: central supermassive black hole, accretion disk, broad line region, dusty molecular torus, narrow line region, and hot X-ray corona. The top view of the diagram also displays a relativistic jet to illustrate that a fraction of AGN possess strong radio emission. 
While these components are typically unresolved by current observational capabilities, recent interferometric studies are achieving unprecedented milliarcsecond resolutions and beginning to spatially resolve outflows and circumnuclear dust structures on scales comparable to the dusty torus in nearby AGN (e.g., \citealt{isbell2023}). Among all the structures of the unified paradigm, the currently most controversial is the torus. Recent observational evidence suggests that it is a structure much more complicated than the simple smooth ``doughnut'' shown in Fig.~\ref{fig:unification} \citep[e.g.,][]{gamez2022,isbell2025}.

The unified paradigm can explain the different properties observed in AGN, particularly the two primary subclasses: unobscured or ``Type I'' AGN and obscured or ``Type II'' AGN. While intermediate classifications exist in the literature, we focus on these two main categories for clarity.

\begin{description}
\item[Type I AGN] exhibit both broad emission lines (FWHM $\gtrsim 1000$\,km\,s$^{-1}$) and narrow emission lines (FWHM $\lesssim 500$\,km\,s$^{-1}$) in their spectra. These objects are viewed at moderate inclination angles where we have a direct view of the broad line region around the central black hole. When such AGN possess radio jets, they are classified as blazars (when viewed along the jet axis) or radio-loud quasars.

\item[Type II AGN] show only narrow emission lines, as the broad line region is obscured by the dusty torus when viewed edge-on. The narrow line region, being more extended, remains visible above the obscuring material. When such AGN possess radio jets, they are classified as radio galaxies.
\end{description}

This unification scheme has been remarkably successful in explaining many observational properties across the electromagnetic spectrum. The presence or absence of broad emission lines becomes a fundamental diagnostic for understanding the geometry and physics of these systems.

\subsection{Emission Line Diagnostics}
\label{subsec:line_diagnostics}

The relative strengths and ratios of forbidden and permitted transitions serve as diagnostics for the physical conditions and ionisation mechanisms in the emitting gas, allowing us to distinguish between AGN and star-formation processes.

\subsubsection{Classical AGN Identification}
\begin{figure}[h]
\sidecaption
\includegraphics[]{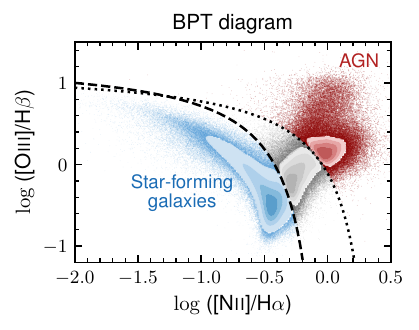}
\caption{BPT diagnostic diagram showing the separation between star-forming galaxies (blue), AGN (red), and composite objects (grey) based on \nii/H$\alpha$ versus \oiii/H$\beta$ line ratios. The loci reflect different ionisation mechanisms: stellar photoionisation produces the star-forming sequence, while AGN show enhanced high-ionisation line ratios from harder continuum spectra. 
Data from \cite{abdurrouf2022}.}
\label{fig:bpt}    
\end{figure}

The most widely used diagnostic scheme employs ratios of strong optical emission lines to distinguish between AGN and star-forming galaxies. The classic ``BPT'' diagram \citep{Baldwin1981} plots the ratio \nii$\lambda$6584/H$\alpha$ versus \oiii$\lambda$5007/H$\beta$. 
These particular lines are strong, while spanning a range of ionisation potentials and critical densities. 
The forbidden lines \oiii\ and \nii\ are collisionally excited and thus probe the electron density and temperature, while the hydrogen recombination lines H$\alpha$ and H$\beta$ trace the ionising photon rate. This two-dimensional parameter space effectively separates three distinct populations (see Fig.~\ref{fig:bpt}):

\begin{itemize}
\item \textbf{Star-forming galaxies}: Occupy a distinct region corresponding to H II region-like ionisation, with line ratios consistent with stellar photoionisation by young, massive stars.
\item \textbf{AGN}: Display enhanced high-ionisation lines due to the harder radiation field (i.e., one with relatively more high-energy photons) from accretion processes.
\item \textbf{Composite objects}: Exhibit signatures of both stellar and AGN photoionisation, typically representing transition cases or spatially mixed systems.
\end{itemize}

\subsubsection{High-Redshift Adaptations}
For sources at higher redshifts, the lines required for the classic BPT diagnostic may be inaccessible within the wavelength coverage of ground-based observatories. Moreover, interpretation becomes more complex, as the loci of these diagnostics appear to shift with redshift, potentially reflecting an evolving mass-metallicity relation and different physical conditions in high-redshift galaxies \citep[e.g.,][]{strom2017}. 

\begin{figure}[h]
\includegraphics[width=\textwidth]{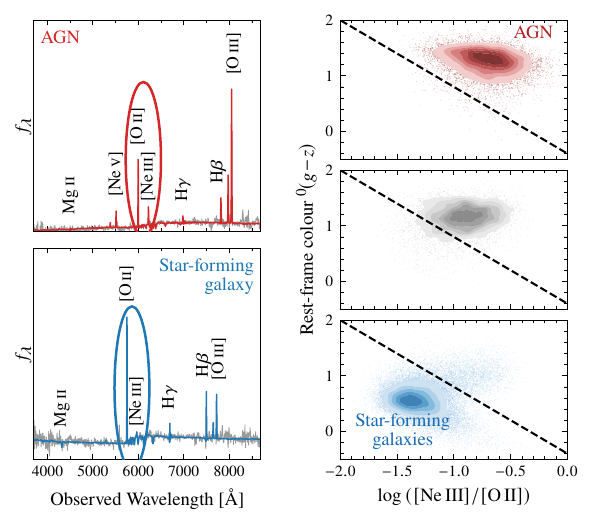}
\caption{Example of a UV-based diagnostic for high-redshift AGN identification using \neiii$\lambda$3869/\oii$\lambda$3727 line ratios. \textbf{Left}: Rest-frame UV spectra of an AGN (top, red) and a star-forming galaxy (bottom, blue), with key emission lines labelled; \neiii\ and \oii\ highlighted (circled). \textbf{Right}: Diagnostic diagram showing AGN (red points) and star-forming galaxies (blue points) separated by \neiii/\oii\ ratio versus rest-frame colour. This method provides effective AGN identification when optical BPT diagnostics are unavailable, and has proven useful up to $z \sim 6$ \citep{backhaus2025}. Spectra from \cite{hviding2024}, diagnostic figure inspired by \cite{trouille2011}.}
\label{fig:neon}    
\end{figure}

Alternative line diagnostics have therefore been developed to address these observational challenges, with new methods continuing to emerge in the JWST era \citep[e.g.,][]{mazzolari2024,trakhtenbrot2025}.   One alternative employs the \neiii$\lambda$3869/\oii$\lambda$3727 combined with rest-frame colour information (see Fig.~\ref{fig:neon}). 
This diagnostic is particularly valuable, as these UV lines remain accessible to ground-based telescopes at higher redshifts than the optical BPT lines. Recent work has demonstrated its continued effectiveness out to $z \sim 6$ with JWST \citep{backhaus2025}. 

The strength of these diagnostics lies in their ability to probe the ionisation state and electron density of the narrow line region, providing insights into the AGN properties even when traditional optical diagnostics become inaccessible or when the broad line region is obscured.

\subsection{The Eddington Limit and Accretion Physics}
\label{subsec:eddington_limit}

A comprehensive treatment of AGN physics is beyond the scope of these lecture notes.  For readers seeking a thorough treatment, I recommend the comprehensive book by \citet{netzer2013}.  Here, I focus on the fundamental physical concepts most relevant to observational studies of high-redshift quasars: the Eddington limit, Eddington luminosity, and Eddington ratio. These concepts appear frequently in the relevant literature; therefore, understanding them is crucial for interpreting AGN observations and for understanding the growth of supermassive black holes over cosmic time. 

\begin{figure}[h]
\sidecaption
\includegraphics[width=0.4\textwidth]{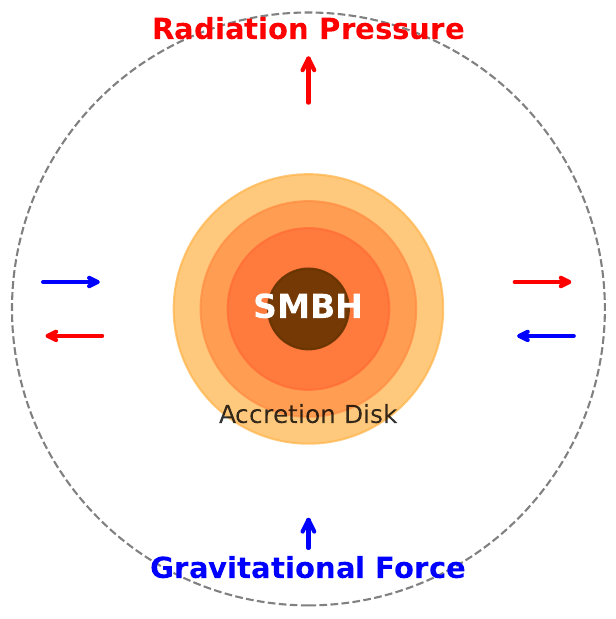}
\caption{Illustration of the Eddington limit.  The balance between outward radiation pressure (red arrows) and inward gravitational force (blue arrows) determines the maximum sustainable luminosity for spherically symmetric accretion onto a supermassive black hole.}
\label{fig:eddington}    
\end{figure}

A fundamental concept governing AGN physics is the \textbf{Eddington limit}, which sets a theoretical maximum luminosity for spherically symmetric accretion onto a compact object. This limit arises from the balance between the gravitational force that pulls matter inward and the radiation pressure that pushes it outward (Fig.~\ref{fig:eddington}).

For a black hole of mass $M_{\rm BH}$, the \textbf{Eddington luminosity}, $L_{\rm Edd}$, is given by:

\begin{equation}
L_{\rm Edd} = \frac{4\pi G M_{\rm BH} m_p c}{\sigma_T} \approx 1.3 \times 10^{38} \left(\frac{M_{\rm BH}}{M_{\odot}}\right) \text{ erg s}^{-1}
\label{eq:eddington_luminosity}
\end{equation}

where $G$ is the gravitational constant, $m_p$ is the proton mass, $c$ is the speed of light, and $\sigma_T$ is the Thomson scattering cross-section. This represents the luminosity at which radiation pressure on electrons (coupled to protons through Coulomb forces) exactly balances gravitational attraction. Exceeding this limit would, in principle, halt further accretion in a spherically symmetric scenario. 

An important parameter for characterising the AGN population is the \textbf{Eddington ratio}: 

\begin{equation}
\lambda_{\rm Edd} = \frac{L_{\rm bol}}{L_{\rm Edd}}
\label{eq:eddington_ratio}
\end{equation}

where $L_{\rm bol}$ is the bolometric luminosity of the AGN. This ratio serves as a fundamental indicator of the accretion state:

\begin{itemize}
\item $\lambda_{\rm Edd} < 0.01$: Low-luminosity AGN, often showing advection-dominated accretion flows
\item $0.01 \lesssim \lambda_{\rm Edd} \lesssim 0.5$: Typical quasar regime with standard thin disk accretion
\item $\lambda_{\rm Edd} \sim 1$: Near-Eddington accretion, common in rapidly growing supermassive black holes
\item $\lambda_{\rm Edd} > 1$: Super-Eddington regime, potentially indicating extreme black hole growth, measurement uncertainties, or non-spherical accretion geometries
\end{itemize}

For context, the supermassive black hole Sagittarius A* in the Milky Way has an extremely low Eddington ratio of $\lambda_{\rm Edd} \approx 10^{-9}$ \citep{EHT2022}, illustrating the quiescent state of our Galaxy's central engine.

\subsubsection{Implications for Black Hole Growth}
The Eddington limit has profound implications for understanding the growth and evolution of supermassive black holes. The maximum accretion rate corresponding to Eddington-limited growth is:

\begin{equation}
\dot{M}_{\rm Edd} = \frac{L_{\rm Edd}}{\epsilon_r c^2} \approx \frac{2.2 \times 10^{-8}}{\epsilon_r} \left(\frac{M_{\rm BH}}{M_{\odot}}\right) M_{\odot} \text{ yr}^{-1}
\label{eq:accretionrate}
\end{equation}

where $\epsilon_r \approx 0.1$ is the typical radiative efficiency for thin disk accretion onto a black hole, representing the fraction of rest mass energy converted to radiation.

This sets a fundamental timescale for black hole growth, known as the Salpeter time, $t_{\rm Salp}$:

\begin{equation}
t_{\rm Salp} = \frac{\epsilon_r c^2}{L_{\rm Edd}/M_{\rm BH}} \approx 450 \left(\frac{\epsilon_r}{0.1}\right) \text{Myr}
\label{eq:eddington_timescale}
\end{equation}

Putting all of this together, an accreting black hole would grow exponentially as:

\begin{equation}
M_{\rm BH}(t) = M_{\rm BH,seed} \times \exp\left[(1-\epsilon_r) \lambda_{\rm Edd} \frac{t}{t_{\rm Salp}}\right]
\label{eq:exponential_growth}
\end{equation}

This rapid growth timescale is crucial for understanding how supermassive black holes with masses $\gtrsim 10^9 M_{\odot}$ can form within the first billion years of cosmic history, as observed in high-redshift quasars \citep{fan2023}.

\subsection{Black Hole Mass Determinations}
\label{sec:bh_mass}
Accurate determinations of supermassive black hole masses are fundamental to understanding the growth and evolution of these objects throughout cosmic history. As we push towards higher redshifts and probe the epoch of the first billion years, measuring black hole masses becomes increasingly challenging. Direct measurements are exceptionally difficult to achieve at high redshift, making indirect methods essential for studying the early universe. Our understanding rests on a hierarchy of measurement techniques, each with distinct strengths, limitations, and systematic uncertainties.

\subsubsection{Direct Black Hole Mass Measurements}
\label{subsec:direct_mass}
The most reliable black hole mass determinations are obtained through direct dynamical measurements that resolve the gravitational sphere of influence—the region where the black hole's gravity dominates over the surrounding stellar potential—via stellar or gas dynamics \citep[e.g.,][]{gillessen2009,ghez2008,barth2016,haberle2024}. These measurements become exceedingly difficult at high redshift due to limited angular resolution and the faint surface brightness of the relevant tracers. In principle, ALMA can achieve the necessary resolution to resolve black hole spheres of influence even at $z>6$, but practical challenges in detecting the required gas dynamics have so far prevented this \citep{meyer2023}.

A remarkable exception is the GRAVITY+ instrument, which has achieved direct black hole mass measurements up to $z \sim 2$ by spatially resolving the broad line region of quasars through interferometric techniques. GRAVITY+ measured 3C~273 at $z = 0.16$ ($\log M_{\rm BH}/M_{\odot} = 8.41 \pm 0.18$; \citealt{sturm2018}), and extended to J0920 at $z = 2$ ($\log M_{\rm BH}/M_{\odot} = 8.51 \pm 0.28$; \citealt{abuter2024}). These measurements, although limited in number, provide empirical calibrations for indirect methods and represent the current gold standard at intermediate redshifts.

\subsubsection{Reverberation Mapping: The Foundation of High-Redshift Studies}
\label{subsec:reverberation}

\begin{figure}[h]
\includegraphics[width=0.99\textwidth]{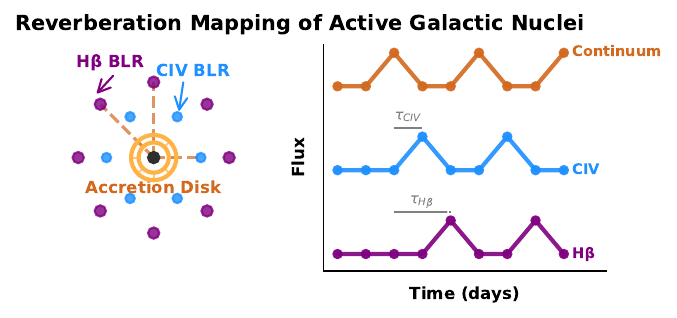}
\caption{Schematic illustration of reverberation mapping in AGN. 
    \textbf{Left:} Physical setup around the central black hole, showing the accretion disk (orange), the inner broad-line region (BLR) traced by \civ\ emission (blue), and the outer BLR traced by H$\beta$ emission (purple). 
    Light-travel time delays between the variable continuum source and the BLR emission lines are indicated with dashed lines. 
    \textbf{Right:} Idealized light curves showing the continuum (orange), \civ\ (blue), and H$\beta$ (purple). 
    The emission-line responses are delayed with respect to continuum variations by characteristic timescales ($\tau_{\rm \civ}$, $\tau_{\rm H\beta}$), reflecting the radial distances of the corresponding BLR regions. 
   These measured delays enable estimation of BLR structure and central black hole mass.}
    \label{fig:reverberation_mapping}
\end{figure}

Reverberation mapping (RM) forms the backbone of black hole mass measurements beyond the reach of direct methods. This technique exploits the light-travel time delay between variations in the central continuum source and the response of the broad emission lines from gas clouds orbiting the black hole (see Fig.~\ref{fig:reverberation_mapping}). 

Sometimes reverberation mapping is referred to as ``light echo'', as the concept is analogous to measuring the size of a cave by timing the echo of a sound—the longer the delay, the larger the cavity. In AGN, the accretion disk around the black hole acts as a variable continuum source, while gas clouds in the broad line region absorb and re-emit this radiation. Changes in the continuum brightness create a signal that propagates outward at the speed of light, causing the broad emission lines to respond after a characteristic time delay $\tau_{\rm lag}$. This time delay directly measures the light-travel time across the BLR:
\begin{equation}
R_{\rm BLR} = c \times \tau_{\rm lag}
\label{eq:blr_radius}
\end{equation}
where $R_{\rm BLR}$ is the characteristic radius of the BLR and $c$ is the speed of light.

Assuming that the gas motion in the BLR is dominated by the gravitational potential of the central black hole and that the system is in virial equilibrium, the black hole mass can be determined from:
\begin{equation}
M_{\rm BH} = \langle f \rangle \frac{R_{\rm BLR} \times \Delta V_{\rm BLR}^2}{G}
\label{eq:virial_mass}
\end{equation}
where $\Delta V_{\rm BLR}$ is the velocity width of the broad emission line (typically measured as the FWHM of the broad emission line), $G$ is the gravitational constant, and $\langle f \rangle$ is a dimensionless virial factor that depends on the unknown geometry and kinematics of the BLR. 
The virial factor $\langle f \rangle$  accounts for uncertainties in the BLR geometry and kinematics, and typically ranges from 1 to 5. This factor has traditionally been calibrated using the $M_{\rm BH} - \sigma_*$ relation in local galaxies \citep[e.g.,][]{yu2019} and more recently through direct measurements like those from GRAVITY+ \citep[e.g.,][]{amorim2020}.

Reverberation mapping has been most successful for nearby AGN ($z \lesssim 0.5$) because their relatively short time delays allow for feasible monitoring campaigns. At higher redshifts, the observed lag between continuum and emission-line variability is stretched by cosmological time dilation by a factor of $(1 + z)$, which lengthens the duration of observations required to detect reverberation signals. These extended timescales make such studies progressively more challenging, often requiring decade-long campaigns \citep[e.g.,][]{lira2018}. 
 These practical limitations have motivated the development of alternative approaches. Continuum reverberation mapping probes the wavelength-dependent light travel time across the accretion disc, providing characteristic timescales that can be $\sim$100 times shorter than  BLR-based measurements \citep[e.g.,][]{panda2024}.  Recently,  
 \cite{pozo2025} reported the first direct measurement of an accretion disc size in a luminous quasar at $z=2.7$, determining a rest-frame ultraviolet-emitting disc radius of $\sim$3 light-days from inter-band continuum time delays.  This result demonstrates that accretion disc reverberation mapping is already feasible at high redshift on timescales of years rather than decades. 
 
The most extensively studied emission line for RM is H$\beta$ $\lambda$4861, which has provided the foundation for black hole mass scaling relations. The H$\beta$ reverberation mapping sample has established robust relationships between BLR size, AGN luminosity, and emission line properties that form the basis for single-epoch mass estimates at higher redshifts.

\begin{important}{Notes on key emission lines in Reverberation Mapping}
Multiple emission lines have been successfully monitored in reverberation mapping campaigns \citep[e.g.,][]{shen2024}. Here I list some key considerations when using these results, ordered by decreasing sample size and reliability:

\begin{itemize}
\item \textbf{H$\beta$\,$\lambda$4861:} The most extensively studied line and current ``gold standard.'' Shows evidence for virialised motion and provides the most reliable mass estimates \citep[e.g.,][]{bentz2009,bentz2013}.

\item \textbf{\civ\,$\lambda$1549:} Well-studied but often affected by non-virial outflows, requiring careful analysis and extensive monitoring to isolate the virialised component \citep[e.g.,][]{lira2018,kaspi2021}.

\item \textbf{\mgii\,$\lambda$2798:} Sample size has grown significantly in recent years, enabling studies at higher redshifts ($z \sim 1.5$) than H$\beta$. Shows good agreement with H$\beta$ results when direct comparisons are available \citep[e.g.,][]{homayouni2020}.

\item \textbf{\halpha\,$\lambda$6563:} The smallest sample, challenging due to the larger BLR radius traced by this line, which produces longer time lags. Current samples are limited to nearby objects ($z \sim 0.01$) with larger scatter compared to other tracers \citep[e.g.,][]{cho2023,dallabonta2025}.
\end{itemize}
\end{important}

\subsubsection{Single-Epoch Mass Estimates: Extending to High Redshift}
\label{subsec:single_epoch}

Reverberation mapping provides the most reliable indirect black hole masses, but the technique is observationally expensive and becomes impractical at high redshifts. Single-epoch (SE) methods have been developed to estimate black hole masses from single spectra by leveraging empirical relationships established from the RM sample.

\begin{figure}[h]
\sidecaption
\includegraphics[width=0.63\textwidth]{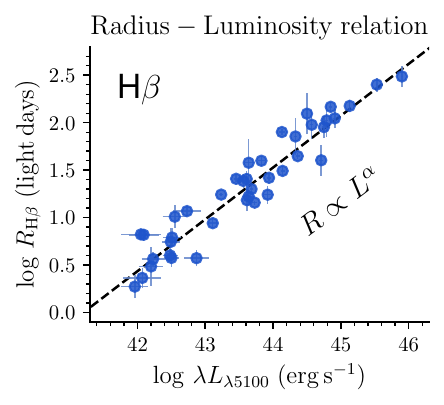}
\caption{Radius--luminosity relation for the H$\beta$ broad emission line,
showing the correlation between the BLR radius (measured via reverberation
mapping time lags) and the AGN continuum luminosity at 5100\,\AA. This correlation enables single-epoch mass estimates, allowing the use of continuum luminosity as a proxy for BLR radius. 
The scatter around the best-fit relation ($\sim$0.2\,dex) represents the
fundamental uncertainty in single-epoch mass estimates when applied to
individual sources.
Data from \citet{bentz2013}.}
\label{fig:RL}     
\end{figure}

The key to SE methods is the empirical \textbf{radius-luminosity relation}, which links the continuum luminosity to the size of the BLR \citep[e.g.,][]{bentz2013}:
\begin{equation}
R_{\rm BLR} = A \left(\frac{\lambda L_\lambda}{10^{44} \text{ erg s}^{-1}}\right)^\alpha
\label{eq:rl_relation}
\end{equation}
where $\lambda L_\lambda$ is the monochromatic luminosity at a specific wavelength (typically 5100\,\AA), and $A$ and $\alpha$ are empirically determined constants. 
For virialized systems, theory predicts $\alpha = 0.5$, and observations of the H$\beta$ line indeed yield $\alpha \approx 0.5$ with some scatter (Fig.~\ref{fig:RL}). 
This agreement provides strong support for the virial interpretation of BLR dynamics.

Combining the radius–luminosity relation (Eq.~\ref{eq:rl_relation}) with the virial mass equation (Eq.~\ref{eq:virial_mass}) yields the basic SE estimator:
\begin{equation}
M_{\rm BH} \propto L_\lambda^\alpha \times {\rm FWHM}^2 ,
\label{eq:se_mass}
\end{equation}
where both the continuum luminosity and the line width (FWHM) can be measured from a single spectrum. 
Because of its practicality, this method underpins nearly all black hole mass estimates at $z>2$, and form the basis of the black hole masses discussed in the following sections (and the other chapters of this book).  
These scaling relations are calibrated against the RM sample \citep[e.g.,][]{vestergaard2006,vestergaard2009,shen2024}.

\begin{important}{Caveats for SE Methods at High Redshift}
While SE methods are indispensable for extending black hole mass measurements to the early universe, their application requires careful attention to several important limitations.

\begin{itemize}

\item \textbf{Calibration:} SE relations are anchored to local and intermediate-redshift RM samples, which may not be representative of high-$z$ populations. Selection effects at high redshift can bias scaling relations away from the calibration set.  

\item \textbf{Possible Evolution:} The radius–luminosity relation itself may evolve with cosmic time, either due to intrinsic changes in AGN physics or to redshift-dependent selection biases.  

\item \textbf{Line Dependence:} Different emission lines (e.g., H$\beta$, \mgii, \civ) trace different regions of the BLR and can introduce systematic offsets. 

\item \textbf{Statistical vs. Individual Accuracy:} Each step in the SE method introduces scatter. Ensemble averages are generally robust, but individual mass estimates can carry large systematic uncertainties.  

\end{itemize}
\end{important}

\subsection{Concluding remarks}
This introductory section provides the foundation for understanding high-redshift black hole observations. The key concepts—AGN physics, mass measurement techniques, and observational signatures—will prove essential as we explore the earliest supermassive black holes discovered before and after JWST.
As in my lectures, I want to conclude by highlighting the key points that I hope this section conveys and that students should retain going forward.  
\begin{important}{Take-Home Messages}
\begin{enumerate}
\item Quasars are the most luminous manifestation of AGN activity.  
\item AGN phases are transient episodes in galaxy evolution.  
\item Their enormous energy output is powered by accretion onto growing supermassive black holes.  
\item Measuring black hole masses is essential but remains one of the major challenges in extragalactic astrophysics.  
\end{enumerate}

\medskip
These fundamentals become increasingly important at high redshift, where observational challenges multiply: spectra become noisier, emission lines shift to less favourable wavelengths, and cosmological effects complicate variability studies. The techniques introduced here represent our primary tools for studying the most massive black holes in the first billion years of cosmic history.

\end{important}

\clearpage
\section{The Pre-JWST Era: Establishing the High-Redshift Quasar Population}
\label{banados-sec:pre-jwst}

In the second of my Saas-Fee lectures, I covered the historical overview of quasars at the redshift frontier, key pre-JWST observational results, and highlighted key open questions. This was mostly a tour through a recent review I wrote with Xiaohui Fan and Rob Simcoe, titled ``Quasars and the Intergalactic Medium at Cosmic Dawn'' \citep{fan2023}. The review focused on unobscured Type-I quasars at $z>5.3$ and studies of the intergalactic medium based on quasar absorption spectra, providing a comprehensive status of the field in the pre-JWST era. This choice was intentional, as we knew JWST would be transformative and that we needed a baseline for comparison before everything changed. 

Rather than repeating that comprehensive material, I encourage readers to consult \cite{fan2023} for the full scope of observational and theoretical developments in high-redshift quasar astronomy before the JWST era. For these written lectures, I will instead summarise those results directly in Section 3, where we will contrast the pre-JWST era with the most recent JWST results.

Therefore, this section will be slightly different than the original lecture. I will update some of the results from \cite{fan2023} with more recent discoveries and results not strictly connected to JWST observations. I will also take the liberty of adopting a more personal approach, describing some of the anecdotes I presented during the lectures, highlighting the collaborative spirit that ultimately prepared our community for the revolutionary capabilities that JWST would bring to our field.

\subsection{The high-redshift quasar landscape}
\label{subsec:quasar-landscape}

Identifying high-redshift quasars is an observationally challenging endeavour. These objects are exceptionally rare, with space densities that decline exponentially with increasing redshift \citep{schindler2023,matsuoka2023}, making them among the rarest objects in the observable universe. This scarcity necessitates large-area, multicolour photometric surveys to identify sufficient candidates for spectroscopic follow-up. 

The selection process is further complicated by numerous contaminants that can mimic the photometric signatures of high-redshift quasars. Survey artefacts—such as moving objects that appear in only one photometric band, cosmic rays, image defects, or spurious detections—can produce dropout signatures similar to those of genuine high-redshift quasars (see, e.g., \citealt{bosman2023}). While these technical contaminants can often be removed through careful quality control and proper motion analysis, astrophysical contaminants present a more formidable challenge.

The most problematic astrophysical contaminants are ultracool dwarfs—low-mass stars and brown dwarfs with spectral types M, L, and T. These objects are point sources with photometric colours that can closely resemble those of high-redshift quasars, particularly in optical and near-infrared colour space. Critically, ultracool dwarfs are far more common than high-redshift quasars by 2--4 orders of magnitude in photometric surveys \citep{barnett2019}. This extreme imbalance means that even highly refined photometric selection techniques must contend with substantial contamination rates; as a result, spectroscopic confirmation is essential but resource-intensive.

\begin{figure}[htbp]
\centering
\includegraphics[width=0.95\textwidth]{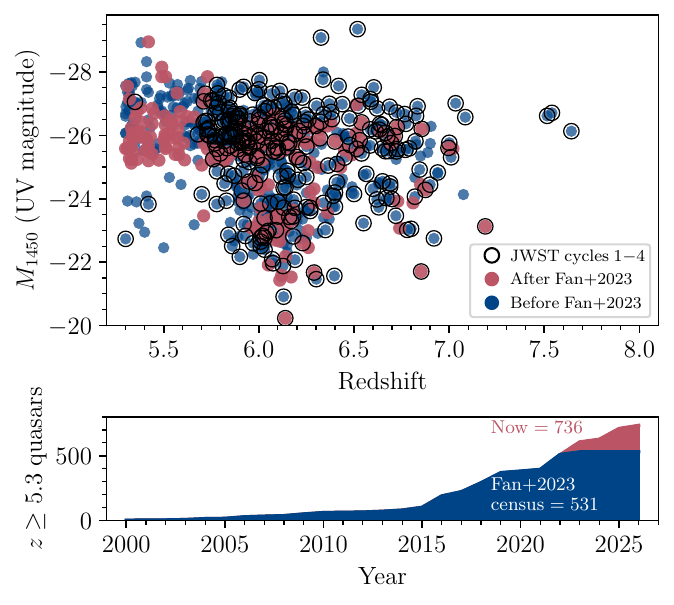}
\caption{Updated census of spectroscopically confirmed quasars at $z \geq 5.3$, extending the compilation presented in \citet{fan2023}. \textit{Top panel:} Distribution in the absolute magnitude-redshift plane. Blue-filled circles show quasars known before the \citet{fan2023} compilation (531 objects), red-filled circles show ground-based discoveries made after that review, and open circles with black edges highlight quasars that have been followed up with JWST observations in Cycles 1--4. 
\textit{Bottom panel:} Cumulative number of $z \geq 5.3$ quasars as a function of publication year. The blue region shows the \citet{fan2023} compilation, while the red region shows subsequent ground-based discoveries.}
\label{fig:quasar-census-update}
\end{figure}

Despite these challenges, systematic surveys have achieved remarkable success in building a census of high-redshift quasars. Figure~\ref{fig:quasar-census-update} presents an updated version of the quasar census originally presented in Fig.~1 of \cite{fan2023}. The top panel shows the distribution of these quasars in the UV absolute magnitude-redshift plane.  The luminosity range spans from the brightest quasars at $M_{1450} \sim -29$ to fainter systems reaching $M_{1450} \sim -22$, thanks particularly to deep surveys like SHELLQs \citep[e.g,][]{matsuoka2022} that have pushed to lower luminosities.

At the time of that review, the total number of spectroscopically confirmed quasars at $z \geq 5.3$ was 531. In over two years since that compilation, the sample has grown to more than 700 objects—almost a 40\% increase. The post-Fan+2023 discoveries (shown in red) demonstrate continued progress, with new quasars being identified across the full redshift and luminosity range covered by earlier surveys. These discoveries come from both the completion of ongoing systematic surveys \citep[e.g.,][]{ighina2024,matsuoka2025ApJS..280...68M,belladitta2025} and new large-area programmes and selection techniques \citep[e.g.,][]{yang2023_desi,byrne2024,yang_daming2024,wolf2024A&A...691A..30W,banados2025MNRAS.542.1088B}, reflecting the continued productivity of the community in completing the quasar census in the first billion years of the Universe. The full updated database is provided as supplementary material accompanying this chapter, serving as an update to the census published alongside the \cite{fan2023} review.

Table~\ref{tab:z7quasars} lists the eleven known quasars at $z \geq 7$, compared to eight in Table~1 of \cite{fan2023}. The three new additions are J1236+6212 at $z = 7.19$ \citep{fujimoto2022,fei2026}, J1330+4234 at $z = 7.02$ \citep{matsuoka2025ApJS..280...68M}, and J0410$-$0139 at $z = 7.00$ \citep{banados2025NatAs...9..293B}. Notably, J1236+6212 (also known as GNz7q) was originally classified as an AGN in its discovery paper and was therefore not included in the \cite{fan2023} quasar census. Recent JWST spectroscopy has since revealed it to be a super-Eddington accreting quasar with $\lambda_{\rm Edd} \approx 2.7$ \citep{fei2026}---a clear example of the classification ambiguities at the faint quasar/AGN boundary ($M_{1450} \sim -23$) discussed in Section~\ref{banados-subsec:agndef} and Fig.~\ref{fig:luminosity}. A notable feature of this sample is the persistent redshift gap between the cluster of eight quasars at $7.00 \leq z \leq 7.19$ and the three most distant quasars at $z \geq 7.5$, with no confirmed quasars in the intervening range. This gap most likely reflects a selection effect and high contamination rates rather than an intrinsic lack of quasars: the $z \geq 7.5$ quasars were discovered by surveys optimised for the very highest redshifts, while systematic coverage at $7.2 \lesssim z \lesssim 7.5$ remains incomplete.

The rate of discovery shows no sign of slowing\footnote{Illustrating this point, during the editorial process of this book I was able to add new discoveries from 
\cite{martinez-ramirez2026} and \cite{davies2026}.} (bottom panel of Fig.~\ref{fig:quasar-census-update}), suggesting that the population of known high-redshift quasars will continue to grow substantially in the coming years with space telescope surveys such as 
\textit{Euclid} \citep{mellier2025} 
and the Nancy Grace Roman Space Telescope \citep[\textit{Roman};][]{schlieder2024}, 
and the ground-based Vera C.\ Rubin Observatory's Legacy Survey of Space and Time \citep[LSST;][]{ivezic2008}.

\begin{table}
\caption{Known quasars at $z \geq 7$, sorted by redshift. 
The full census of $z > 5.3$ quasars presented in this chapter
is publicly available at \citet{banados2026_catalog}
and will be updated regularly. 
\label{tab:z7quasars}}
\begin{tabular}{lccll}
\hline\hline
Quasar & $z$ & $M_{1450}$ & Disc.\ ref. & $z$ ref. \\
\hline
J031343.84-180636.40 & 7.64 & $-26.13$ & \cite{Wang2021ApJ...907L...1W} & \cite{Wang2021ApJ...907L...1W} \\
J134208.11+092838.61 & 7.54 & $-26.71$ & \cite{banados2018a} & \cite{banados2019a} \\
J100758.27+211529.21 & 7.51 & $-26.62$ & \cite{yang2020} & \cite{yang2020} \\
J123616.92+621232.13 & 7.19 & $-23.13$ & \cite{fujimoto2022} & \cite{fujimoto2022} \\
J112001.48+064124.30 & 7.08 & $-26.58$ & \cite{mortlock2011} & \cite{2020ApJ...904..130V} \\
J124353.93+010038.50 & 7.07 & $-24.13$ & \cite{matsuoka2019} & \cite{izumi2021ApJ...914...36I} \\
J003836.10-152723.60 & 7.03 & $-27.01$ & \cite{wang-feige2018ApJ...869L...9W} & \cite{yang2021} \\
J133013.00+423432.20 & 7.02 & $-25.58$ & \cite{matsuoka2025ApJS..280...68M} & \cite{matsuoka2025ApJS..280...68M} \\
J235646.33+001747.30 & 7.01 & $-25.31$ & \cite{Matsuoka2019ApJ...883..183M} & \cite{Matsuoka2019ApJ...883..183M} \\
J025216.64-050331.80 & 7.00 & $-25.77$ & \cite{yang2019b} & \cite{wang2024ApJ...968....9W} \\
J041009.05-013919.88 & 7.00 & $-25.61$ & \cite{banados2025NatAs...9..293B} & \cite{banados2024} \\
\hline
\end{tabular}
\end{table}

\subsection{Quasars at the Redshift Frontier: $z > 7$}

\begin{figure}[htbp]
\centering
\includegraphics[width=0.95\textwidth]{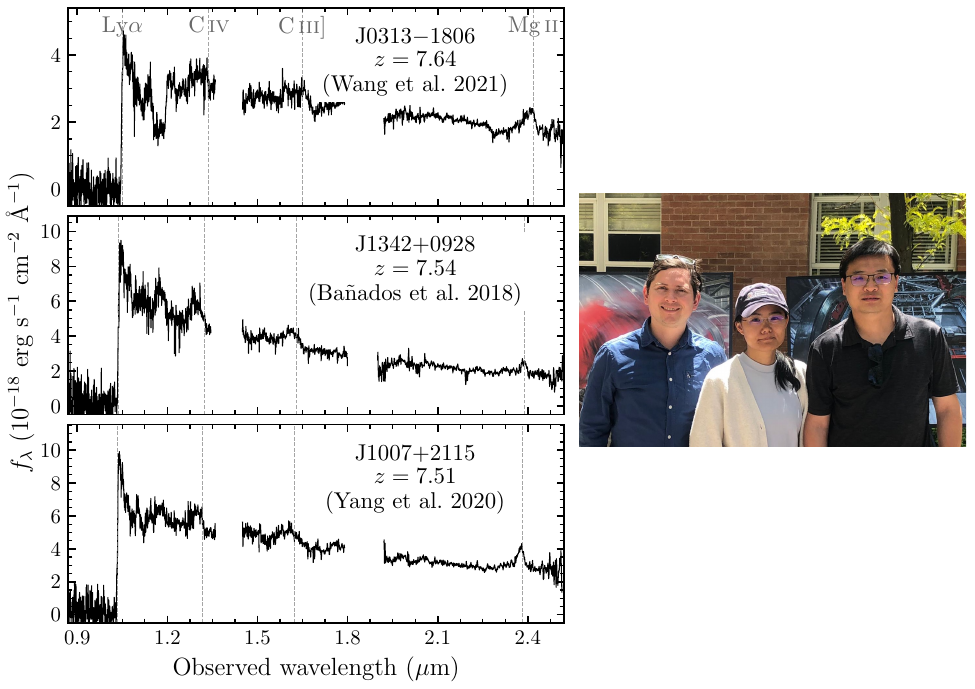}
\caption{The three most distant quasars currently known and their lead discoverers. 
 \textbf{Left}: Ground-based near-infrared spectra of the three most distant quasars, showing the characteristic features including complete Gunn-Peterson absorption blueward of the Ly$\alpha$ emission line (which falls at $\sim$1$\mu$m at these redshifts).  \textbf{Right}: The three lead authors of these discoveries: Eduardo Bañados (left), Jinyi Yang (middle) and Feige Wang (right). Photo taken during the 20th anniversary of the Large Binocular Telescope in April 2024.}
\label{fig:z7frontier}
\end{figure}

At the time of the \citet{fan2023} 
review (late 2022), 
eight quasars were known at $z > 7$, with three beyond $z=7.5$: J0313$-$1806 at $z = 7.64$ \citep{Wang2021ApJ...907L...1W}, J1342+0928 (Pisco) at $z = 7.54$ \citep{banados2018a}, 
and J1007+2115 (Pōniuā\textquoteleft{}ena) at $z = 7.51$ \citep{yang2020}. 
These three objects remain the most distant luminous quasars known as of this writing. They have provided some of the strongest constraints on supermassive black hole growth, build-up of massive galaxies, large scale structure, and the state of the intergalactic medium when the universe was less than 700 Myr old—barely 5\% of its current age (see Fig.~7 of \citealt{fan2023}). 
Each of these discoveries represents a remarkable achievement. Figure~\ref{fig:z7frontier} shows the ground-based spectra of these three quasars alongside a group photo of their lead discoverers: Feige Wang, myself, and Jinyi Yang.

Beyond simply adding objects to the census, this small sample of $z > 7$ quasars has enabled key scientific advances. In the following subsections, I explore three examples: the discovery story of Pisco (\S\ref{subsec:pisco}), how these quasars probe cosmic reionization through IGM damping wings (\S\ref{subsec:reionization}), and what a recently discovered $z=7$ blazar reveals about the completeness of our census (\S\ref{subsec:blazar}).

\subsubsection{The Pisco Discovery}
\label{subsec:pisco}

From the outside, each quasar in Figure~\ref{fig:quasar-census-update} may appear as just another data point. But these discoveries are made by people, and rarely by chance. 
Many of these quasars have their own backstories, their own moments of excitement and challenge. Here, as in my original Saas-Fee lecture, I would like to share the behind-the-scenes story of how we discovered J1342+0928—the ``Pisco quasar"—to illustrate what these discoveries really entail, and to explain why we gave it this particular name.

During my first year of postdoc as a Carnegie-Princeton Fellow at Carnegie Observatories, I had the privilege of having regular access to the twin 6.5\,m Magellan Telescopes. This access, alongside the main goals of my observing programmes, gave me time to explore the capabilities of the instruments. By mid-2016, I had demonstrated that the FIRE spectrograph \citep{simcoe2008} could identify and confirm a $z=7.1$ quasar---then the most distant known \citep{mortlock2011}---in under 5 minutes of exposure time. 

This efficiency was key: with FIRE, I could rapidly screen multiple candidates per night. Thus, I wrote a proposal titled \textit{The Carnegie search for the first QSOs: pushing the redshift barrier}, arguing that a systematic spectroscopic follow-up of carefully selected candidates could discover quasars beyond $z=7.1$.  The time allocation committee awarded me three nights in March 2017. I spent months systematically compiling candidates---sources that were bright in near- and mid-infrared surveys like UKIDSS and WISE but invisible in optical surveys, as expected for objects at $z>7.1$.

In February 2017, while preparing for this observing run, I hosted a workshop at Carnegie (which I will describe in Section~\ref{sec:jwst_prep}), where I presented this upcoming programme. I also asked whether anyone would like to join me as co-observer for this intensive campaign. My collaborator Dan Stern stepped up.  I was happy to have Dan on board: a few months earlier he had invited me to observe with him at Palomar Observatory,  and I had learnt a great deal from that experience. 

When we arrived at Las Campanas Observatory, Dan and I immediately confronted a practical challenge: we had far too many candidates, and I was concerned that many might not even be real astronomical sources. They could be artifacts, cosmic rays, or moving objects. We decided to devote the first night to rapid imaging of as many candidates as possible using the FourStar near-infrared camera \citep{persson2013}.  This would confirm which sources were real and whether their photometry matched expectations from the large-area surveys. This ``point and shoot'' strategy proved highly efficient at eliminating spurious detections, allowing us to focus our precious spectroscopic time on only the most promising candidates. 

The second night brought disappointment: poor observing conditions kept the dome closed for much of the night. This gave us time to reduce and analyse the FourStar imaging data from night one, but the anxiety was real. With only one night remaining, we worried whether the weather would cooperate or whether we might return empty-handed. 

On the third night, conditions were marginal. Not optimal, but good enough to begin spectroscopy. We were in the zone: I drove the telescope as efficiently as possible while Dan, in parallel, reduced and analysed each spectrum as it came off the detector, searching for the telltale signatures of a distant accreting supermassive black hole. 

After 3 am, we still hadn't found a single quasar. Our morale began to fade as the night wore on and target after target turned out to be contaminants. Then, at about 5 am, Dan called out: ``Come and look—we might have something here."

\begin{figure}[htbp]
\centering
\includegraphics[width=0.95\textwidth]{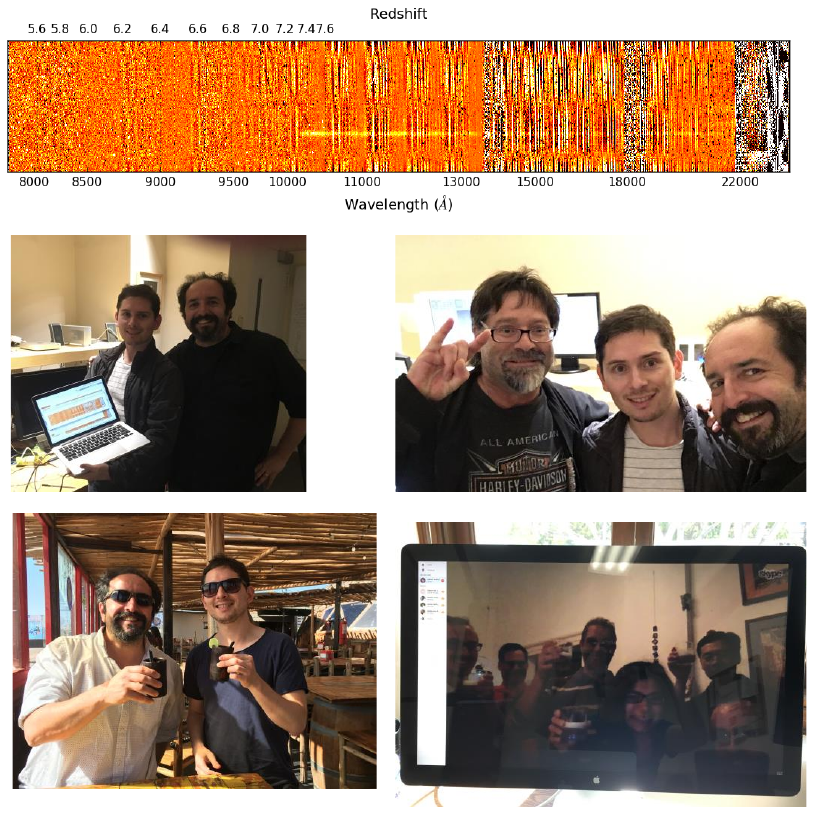}
\caption{Discovery and celebration of the $z=7.54$ Pisco quasar (J1342+0928). \textbf{Top:} Initial discovery spectrum, obtained 09 March 2017. This raw spectrum never appeared in the formal publication \citep{banados2018a}. \textbf{Middle:} Moments after the 5 am discovery at Las Campanas Observatory. \textit{Left}: Dan Stern and Eduardo Bañados with the spectrum in the control room. \textit{Right}: Celebrating with telescope operator Jorge Araya (left). \textbf{Bottom:} Toasting with Pisco Sours. \textit{Left}: At the beach near La Serena after the observing run. \textit{Right}: Video call with the team in Heidelberg, Germany—from left: Roberto Decarli, Fabian Walter, Bram Venemans, Chiara Mazzucchelli, Xiaohui Fan, and Emanuele Farina.}
\label{fig:pisco-discovery}
\end{figure}

One glance at the spectrum told us we'd hit a winner. But we needed to be certain. We slewed the telescope back to the source and took a deeper 10-minute exposure to confirm what we were seeing. The signal was unmistakable: a strong Ly$\alpha$ break indicating $z \gtrsim 7.2$, accompanied by the characteristic UV spectral slope of a luminous quasar. Figure~\ref{fig:pisco-discovery} shows this initial discovery spectrum—which never appeared in our formal publication—alongside photographs taken just after we realised what we had found. We were confident we had discovered a redshift record-breaking quasar, though precise characterisation of the redshift and emission lines would require additional observations and a more careful data reduction \citep{banados2018a}. 

Sleep was impossible. Before closing the telescope, I sent an email to the team with the good news. The subject line read: \textit{pisco sour quasar}. Pisco Sour is a typical celebratory cocktail and traditional drink in Chile and Peru. Right after that final observing night, Dan and I headed to the beach near the city of La Serena to toast our discovery with a Pisco Sour.  When I returned to my office in Pasadena a few days later, I had a video call with the team in Germany. They too raised Pisco Sours in honour of the discovery.  Since then, we have affectionately referred to quasar J1342+0928 as ``Pisco''.

\subsubsection{Reionization Constraints from $z>7$ Quasars}
\label{subsec:reionization}

The most distant quasars not only probe galaxy and black hole growth but also serve as unique tools to study the state of the intergalactic medium (IGM) during the epoch of reionization. Section 6 of \cite{fan2023} provides a thorough review of how quasars can be used to constrain reionization, with their Figure~16 summarising the observational constraints from various techniques.

Several complementary methods have been developed to probe the reionization history using quasar absorption spectra. These include measurements of the Ly$\alpha$ forest optical depth, statistics of dark gaps and transmission spikes, the mean free path of ionising photons, and, most relevant to $z > 7$ quasars, the detection of IGM damping wing absorption. 

\begin{important}{Key Constraints on Reionization from High-Redshift Quasars}
Quasar absorption spectra provide our most direct probes of the ionisation state of the IGM during the epoch of reionization. The main conclusions are:
\begin{itemize}
\item Reionization was essentially complete by $z \sim 5.3$, evidenced by Ly$\alpha$ forest transmission and the absence of saturated Gunn-Peterson troughs \citep[e.g.,][]{fan2006c,bosman2022}.
\item Rapid evolution in IGM ionisation occurs at $z > 6$, marked by saturated Gunn-Peterson troughs, increasingly long dark gaps, and a steep decline in the mean free path of ionising photons \citep[e.g.,][]{mcgreer2015,jin2023,durovcikova2024,greig2024,davies2025}.
\item At $z \sim 7$--7.5, the volume-averaged neutral fraction reached $\bar{x}_{\rm HI} \sim 0.3$--0.7, constrained by IGM damping wing detections in $z > 7$ quasar spectra \citep[e.g.,][]{davies2018b,greig2019,greig2022,durovcikova2022}.
\end{itemize}
Combined with Cosmic Microwave Background measurements and high-redshift galaxy observations, these constraints indicate that reionization had its midpoint at $z\sim 7-8$, was largely complete by $z\sim 6$, with residual neutral gas fluctuations persisting until $z\sim 5.3$. 
\end{important}

I will not repeat the comprehensive technical details here, as they are covered in \cite{fan2023}. Instead, I want to share how I became aware of IGM damping wings—a story that connects directly to one of my fellow Saas-Fee lecturers.

\runinhead{Learning about IGM damping wings.} 
The IGM damping wing is a long-predicted signature for sources embedded in a significantly neutral IGM. When a quasar resides in an environment with neutral fractions exceeding $\sim$$10\%$, the effects of Ly$\alpha$ resonance scattering become so significant that one can detect characteristic absorption in the spectrum \textit{redward} of the Ly$\alpha$ emission line—a region that would otherwise show unabsorbed quasar continuum. This phenomenon was elegantly described in the paper by \cite{miralda-escude1998}, though it would take nearly two decades before strong observational signatures of this predicted process were detected.

Given that I completed my PhD working on high-redshift quasars, it is somewhat embarrassing to admit that IGM damping wings were not on my radar until I met one of the other lecturers of this Saas-Fee school: Richard Ellis. I confirmed during the school that Richard has no recollection of this conversation, but it left a lasting mark on me.

It was 2 December 2014. I was on a ``job tour" as a finishing PhD student applying for postdoc positions. I visited Pasadena, where I had applied for positions at both Caltech and the Carnegie Observatories, and gave talks at both institutions. Richard Ellis was at Caltech at the time, and though he couldn't attend my seminar, I managed to secure a one-to-one appointment to summarise my thesis work for him.

After my summary, Richard asked: ``\textit{Have you checked whether you've detected any IGM damping wing in your quasar spectra?}''

The question caught me off guard. I did not know what he was referring to and had to admit so. Richard then stood up, walked to the blackboard, and patiently explained the concept. He sketched out how the damping wing absorption would appear in a quasar spectrum embedded in a partially neutral IGM, and how this differed from the on-resonance Gunn-Peterson absorption that saturates at much lower neutral fractions. He also pointed me to the \cite{miralda-escude1998} paper. I wish I had taken a photograph of that blackboard. 

\begin{figure}[h]
\centering
\includegraphics[width=0.675\textwidth]{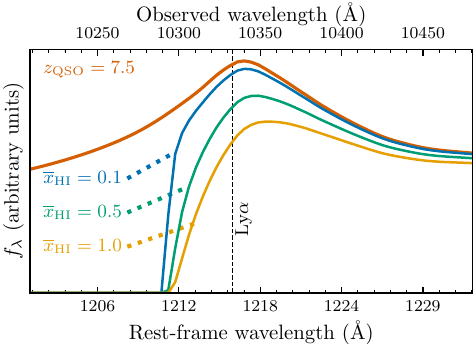}
\caption{Illustration of IGM damping wing signatures in high-redshift quasar spectra. Synthetic spectra of a quasar at $z_{\rm QSO} = 7.5$ for different IGM neutral fractions. The \textit{top spectrum} (orange) shows the intrinsic quasar emission around the Ly$\alpha$ line without IGM absorption. The solid curves below show observed spectra with progressively higher neutral fractions: $\bar{x}_{\rm HI} = 0.1$ (blue), $\bar{x}_{\rm HI} = 0.5$ (green), and $\bar{x}_{\rm HI} = 1.0$ (yellow). The vertical dashed line marks systemic Ly$\alpha$. As the neutral fraction increases, damping wing absorption becomes increasingly prominent redward of Ly$\alpha$. 
Note that the fluz drop to zero does not always occur at $\sim$1212\,\AA: it sifts to shorter (longer) wavelengths for larger (smaller) ionised proximity zones, which vary between quasars \citep[e.g.,][]{eilers2017a}.}
\label{fig:damping-wing-models}
\end{figure}

Since then, whenever I explain IGM damping wings in presentations, I've attempted to reproduce how Richard explained the concept to me at that blackboard. Figure~\ref{fig:damping-wing-models} shows my version of this explanation, illustrating how the observed spectrum of a $z = 7.5$ quasar changes depending on the neutral fraction of the surrounding IGM.

\runinhead{From question to answer.} 
Richard Ellis's question stayed with me, though it took some time before it led anywhere concrete. My initial attempts to look for damping wings in high-redshift quasar spectra led me to identify a first candidate, quasar P183+05 at $z=6.4$ \citep{mazzucchelli2017b}. The damping wing signature was so strong that it appeared suspicious, since many known quasars at similar redshift did not show such a signal.  After further inspection of its NIR spectrum we concluded that this was not the IGM damping wing signature I had been looking for. Instead, this peculiar profile was produced by an intervening close absorption system revealed by several metal absorption lines. The analysis of this absorption system was eventually published in \citet{banados2019b}, though the publication was delayed because the Pisco discovery occurred while I was writing it up. 

One week after the discovery of Pisco, back in my office, I sent the following email to my mentors Xiaohui Fan, Fabian Walter, and Bram Venemans: 

\smallskip
\noindent \textit{This is the combined spectrum, and probably our final FIRE spectrum for the paper. It shows a clear damping wing and I can't identify metals at a similar redshift}.
\smallskip

\begin{figure}[htbp]
\centering
\includegraphics[width=0.85\textwidth]{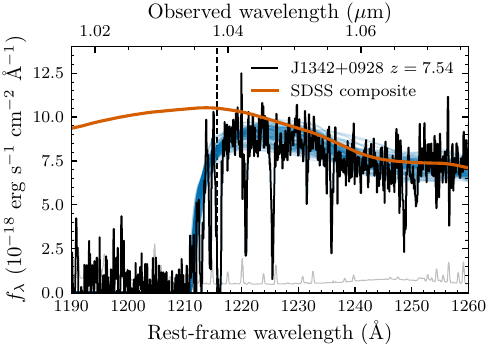}
\caption{IGM damping wing absorption in J1342+0928 (the Pisco quasar) at $z = 7.54$. The observed spectrum (black) shows absorption extending redward of the Ly$\alpha$ emission line peak (vertical dashed line). The red curve shows the estimated intrinsic quasar continuum based on lower-redshift SDSS quasars.  The coloured blue curves show model predictions for different IGM neutral fractions, favouring an intermediate neutral fraction of $\bar{x}_{\rm HI} = 0.4-0.6$. This indicates that the universe was undergoing rapid reionization at $z \sim 7.5$. Figure adapted from \citet{banados2018a}.}
\label{fig:pisco-damping}
\end{figure}

Figure~\ref{fig:pisco-damping} shows the spectrum of J1342+0928 with the damping wing clearly visible as suppressed flux extending redward of the Ly$\alpha$ emission line. Detailed modelling of this absorption profile suggested an IGM neutral fraction of $\bar{x}_{\rm HI} \sim 0.4$--0.6 at $z \sim 7.5$ \citep{banados2018a,davies2018b,greig2019}.

In 2017 I was invited to present my work at a Lorentz workshop in Leiden, one of the first places where I showed Pisco to the world. During the discussion, Richard Ellis asked if we saw any evidence of a damping wing on this $z=7.54$ quasar. This time I knew what he was talking about and I had the answer he wanted to hear. Richard's question had come full circle. 

The damping wing detection in J1342+0928, combined with similar signatures in other $z > 7$ quasars, has provided some of our strongest constraints on the late stages of reionization \citep{davies2018b,Wang2020ApJ...896...23W}. These measurements suggest that at $z \sim 7$--7.5, the universe was transitioning from a largely neutral to a largely ionised state, with volume-averaged neutral fractions of order 30--70\%. The spatial fluctuations in ionisation during this epoch appear to have been substantial. Individual sightlines show significant variations depending on their local environments and proximity to ionising sources \citep{kist2025}.

\subsubsection{Blazars Beyond $z=6$: A Census Puzzle}
\label{subsec:blazar}

Beyond quasars as probes of the IGM, individual discoveries can reveal unexpected insights about the broader population. 
Blazars—quasars with relativistic jets pointed nearly along our line of sight—are particularly informative because geometric arguments allow us to infer the total population of jetted sources from even a single detection.

The first blazar beyond $z=6$ was discovered at $z=6.1$ \citep{belladitta2020}. This detection motivated intensive radio searches for additional jetted quasars, leading to rapid expansion of the census in just a couple of years \citep[e.g.,][]{banados2021,gloudemans2022,ighina2024,belladitta2025}. More recently, a blazar was identified at $z=7.0$ \citep{banados2025NatAs...9..293B}, pushing to even higher redshifts. Unlike the $z=6.1$ discovery, this $z=7$ blazar creates a striking tension with our understanding of the quasar population at the highest redshifts.

\runinhead{The geometric argument.} When we observe a blazar, we are viewing a relativistic jet oriented within a small angle $\theta \lesssim 1/\Gamma$ of our line of sight, where $\Gamma$ is the bulk Lorentz factor of the jet. Based on simple geometric considerations \citep{volonteri2011,belladitta2020}, the existence of this blazar implies many more intrinsically similar jetted sources must exist with jets pointing elsewhere. The number of blazars with viewing angle $\theta\leq 1/\Gamma$ is related to the total jetted population as:
\begin{eqnarray}
    N(\theta\leq1/\Gamma) & = & N_{\rm jetted} \times \dfrac{2\Omega}{4\pi} \nonumber\\
     & = & N_{\rm jetted} \int^{1/\Gamma}_0 \,\sin(\theta)\,d\theta \nonumber\\
     & = & N_{\rm jetted} \left[1 - \cos\left(\frac{1}{\Gamma}\right)\right]
\end{eqnarray}
where $\Omega$ is the solid angle subtended by the jet. Using the Taylor expansion of $\cos(1/\Gamma)$, this simplifies to:
\begin{equation}
N_{\rm jetted} \approx N_{\rm blazar} \times 2\Gamma^2
\end{equation}
For a typical Lorentz factor of $\Gamma\sim10$, this implies $N_{\rm jetted} \approx 2\times 10^2=200$ jetted quasars should exist for every blazar, all with similar intrinsic properties (e.g., black hole masses and redshifts).  If the fraction of quasars hosting relativistic jets is known, then the census of the total population of quasars can be estimated as: 

\begin{equation}
N_{\rm Total} = \frac{N_{\rm jetted}}{\rm Jetted~fraction}
\label{eq:ntotal}
\end{equation}

\begin{figure}[htbp]
\centering
\includegraphics[width=0.85\textwidth]{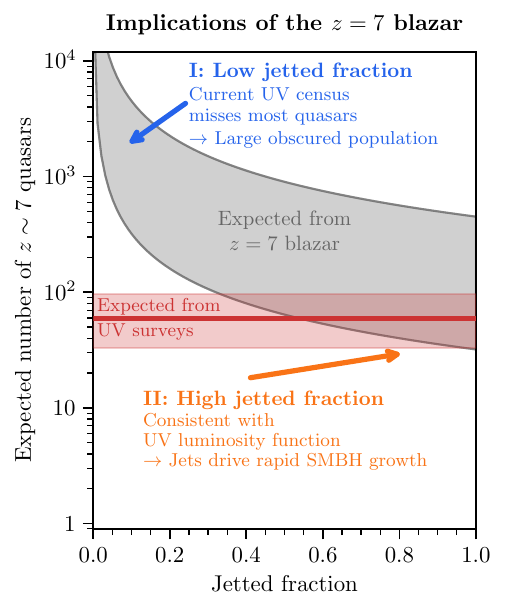}
\caption{{Implications of the $z=7$ blazar discovery.} Grey region: expected number of $z\sim7$ quasars implied by a single blazar detection \citep{banados2025NatAs...9..293B}. Red band: UV luminosity function prediction \citep{matsuoka2023}. Two scenarios reconcile these: (I)~low jetted fraction implies a large obscured population; (II)~high jetted fraction implies jets are ubiquitous and drive rapid SMBH growth.}
\label{fig:blazar-implications}
\end{figure}

\runinhead{The $z=7$ tension.} Applying this geometric argument to the $z=7$ blazar reveals a striking discrepancy. The expected number of jetted quasars at $z\sim7$, and consequently the total quasar population implied by Equation~\ref{eq:ntotal}, significantly exceeds the number predicted by UV luminosity functions \citep{matsuoka2023}. This tension is illustrated in Figure~\ref{fig:blazar-implications}. Two scenarios can reconcile this discrepancy, each with transformative implications for our understanding of early black hole growth:

\textit{Scenario I: Most high-redshift quasars are obscured.} If the jetted fraction is low, the blazar discovery implies that UV-selected surveys miss the majority of $z > 7$ quasars. Such an obscured phase may be essential for rapid growth, as it enables sustained super-Eddington accretion \citep{johnson2022}.

\textit{Scenario II: Jets are ubiquitous and drive rapid SMBH growth.} If most high-redshift quasars host jets, the blazar expectation is consistent with current UV surveys. In this case, jet-enhanced accretion \citep{jolley2008,connor2024} may be the dominant mechanism enabling SMBHs to reach billion-solar masses within 700\,Myr, potentially allowing smaller seeds (e.g., Population~III star remnants) to produce the observed $z > 6$ quasar population.

These scenarios are not mutually exclusive. Distinguishing between them requires building a complete census of $z > 7$ quasars through multi-wavelength surveys combining UV/optical selection with deep radio observations to identify both obscured sources and those with jets. The rapid progress following the $z=6.1$ blazar discovery demonstrates the power of this approach, while the $z=7$ blazar highlights fundamental gaps in our understanding that remain to be addressed.

\begin{figure}[htbp]
\centering
\includegraphics[width=0.99\textwidth]{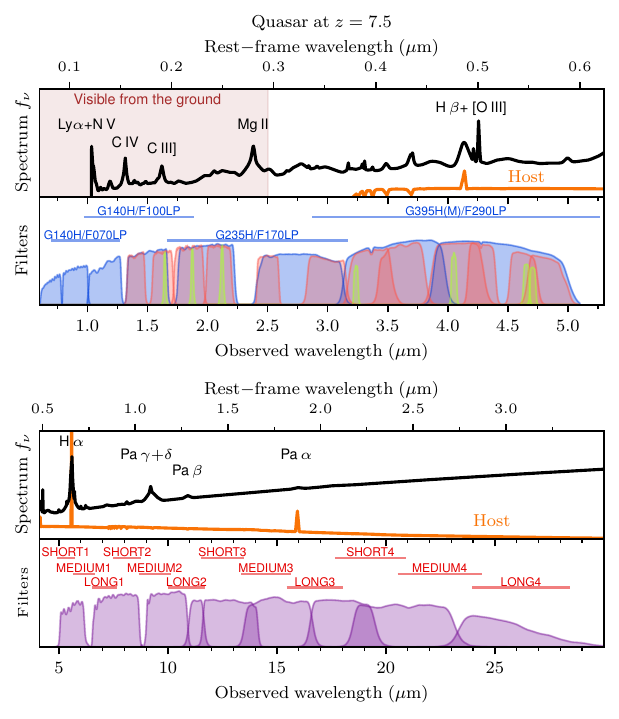}
\caption{Wavelength coverage provided by JWST for a $z = 7.5$ quasar (compare with Fig.~\ref{fig:z7frontier}). \textbf{Top}: Observed wavelengths 1--5\,$\mu$m showing key emission lines accessible with NIRSpec spectroscopy. Black spectrum shows the quasar, orange shows the host galaxy scaled $\times$30. Coloured bands indicate NIRCam filter coverage, while horizontal lines show NIRSpec grating configurations (G140H, G235H, G395H). Red box highlights wavelengths observable from the ground. \textbf{Bottom}: Observed wavelengths 5--28\,$\mu$m showing hydrogen Paschen and Balmer lines, accessible with MIRI spectroscopy. Red and purple bands show MIRI grating and filter coverage. This wavelength range, impossible to achieve from the ground, enables comprehensive characterisation of both AGN and host galaxy properties.}
\label{fig:jwst_capabilities}
\end{figure}

\subsection{Getting Ready for JWST}
\label{sec:jwst_prep}
The years leading up to JWST's launch in December 2021 were 
marked by intensive community preparation. 
We knew that JWST would be transformative for high-redshift quasar science,
but translating that potential into reality required careful planning and coordination.

In 2017, I organised a small workshop at the Carnegie Observatories, ``High-$z$ quasars in the JWST era.'' The workshop format balanced presentations reviewing the state of the field with extensive discussions focused on defining key science questions and developing observational strategies. It was at this meeting that I first met Marta Volonteri, one of my fellow Saas-Fee lecturers, whose theoretical work on black hole formation and growth provided crucial context for interpreting what JWST might reveal.

As the launch date approached, I organised a second, more focused workshop in 2020 to prepare Cycle 1 observing proposals. This meeting proved quite successful, helping to coordinate community efforts and identify the most compelling targets and observing strategies. The expectations were high: JWST's capabilities would enable observations simply impossible from the ground, including rest-frame optical spectroscopy revealing emission lines from ionised gas and stellar populations in quasar host galaxies, as well as mid-infrared coverage accessing the Balmer lines and probing warm dust emission (Figure~\ref{fig:jwst_capabilities}).

The quasar community has been remarkably active since JWST began operations. Figure~\ref{fig:quasar-census-update} shows the updated census of spectroscopically confirmed quasars at $z \geq 5.3$ in the epoch of reionization. JWST has rapidly expanded our ability to characterise these sources in detail. Over the first four observing cycles, hundreds of these quasars have been targeted with JWST, spanning basically the full suite of its capabilities. In the next section, we turn to some of the first emerging results from these observations.

\subsection{Concluding Remarks}

This section has traced the development of high-redshift quasar astronomy in the pre-JWST era, from the systematic surveys that built the census to individual discoveries that revealed unexpected insights about the early Universe. As in my lectures, I have chosen to highlight the human side of this science: the collaborations, the chance encounters, the moments of discovery that are as much a part of these results as the data themselves.

The pre-JWST census of $z \geq 5.3$ quasars grew to over 700 objects through systematic ground-based efforts spanning two decades. Individual objects like the Pisco quasar provided our strongest constraints on cosmic reionization, whilst the discovery of blazars beyond $z=6$ forced us to confront fundamental questions about the completeness of our census. The intensive community preparation for JWST, exemplified by the workshops described in Section~\ref{sec:jwst_prep}, created the foundation for the transformative science that would follow.

\begin{important}{Take-Home Messages}
\begin{enumerate}
\item The census of $z \geq 5.3$ quasars has grown from 531 objects (late 2022) to 700. 

\item The most distant quasars demonstrate that billion-solar-mass black holes and their massive host galaxies existed when the Universe was less than 700\,Myr old.

\item IGM damping wing absorption in $z > 7$ quasar spectra provides direct evidence for substantial neutral hydrogen ($\bar{x}_{\rm HI} \sim 0.3$--0.7) at $z \sim 7$--7.5.

\item The discovery of blazars at $z > 6$ creates a census puzzle: geometric arguments suggest either (a) UV-selected surveys miss the majority of high-redshift quasars due to obscuration, or (b) most early quasars host relativistic jets.

\item Community preparation through workshops and coordinated proposal efforts meant the quasar community was ready to use JWST effectively from its first observing cycles.

\end{enumerate}
\medskip
These pre-JWST foundations provide the baseline against which we can measure JWST's impact, the subject of Section~\ref{banados-sec:jwst}. 
\end{important}

\section{The JWST Revolution}
\label{banados-sec:jwst}

JWST's impact on high-redshift quasar science differs markedly from its transformative role in galaxy discoveries. The limited survey areas accessible to JWST are insufficient for discovering rare, luminous quasars—no new luminous quasars have been found by JWST at $z>6$. However, the observatory excels at detailed follow-up, offering unprecedented characterisation of quasars discovered by wide-field surveys.

Before proceeding, it is worth emphasising what this section does \textit{not} cover. As mentioned in Section~\ref{banados-sec:pre-jwst}, the pre-JWST status of reionization-era quasars was documented in the \cite{fan2023} review; we will not repeat that material but instead reference it where appropriate and provide a summary for context. The reionization constraints from quasar absorption spectra, including the IGM damping wing measurements that probe the neutral fraction, were addressed in Section~\ref{banados-sec:pre-jwst}, but we note that JWST has also started improving these measurements. The higher signal-to-noise and broader wavelength coverage has motivated the development of improved damping wing models \citep{kist2025,kist2026}. Jetted radio quasars, while briefly mentioned in Section~\ref{subsec:blazar}, are not a focus here, as JWST results on these sources are yet to come. And the broader JWST galaxy population---Little Red Dots (LRDs), lower-luminosity AGN candidates, and the census of star-forming systems---is addressed in other lectures at this school. Our focus is specifically on what JWST has revealed about the luminous quasar population and its immediate surroundings.

Following the structure of my lectures, this section is organised by physical scale. For each scale, we first establish the pre-JWST baseline---what was known, what methods were available, and what questions remained open. We then examine how JWST observations have transformed that baseline, identifying both confirmations of earlier work and genuine surprises.


\subsection{Parsec Scales: The Central Engine}
\label{subsec:parsec}

One of the most striking results from JWST observations of high-redshift quasars is how \textit{familiar} the central engines appear \citep{bosman2024,bosman2026}. 
Despite residing in a universe less than a billion years old, the supermassive black holes powering $z>6$ quasars show spectral properties remarkably similar to their lower-redshift counterparts. This continuity across cosmic time validates decades of ground-based work while simultaneously deepening the puzzle of how such massive objects formed so quickly.

\runinhead{Pre-JWST: The Mg\,{\sc ii} Era.}
As discussed in Section~\ref{banados-sec:intro}, measuring black hole masses at high redshift relies on single-epoch virial estimators calibrated against reverberation mapping samples. Before JWST, the Mg\,{\sc ii}$\lambda$2798 emission line served as the workhorse for $z>6$ black-hole mass measurements. At these redshifts, Mg\,{\sc ii} shifts into the near-infrared $K$-band ($\sim$2.0--2.4\,$\mu$m for $z\sim6$--7.5), where ground-based spectroscopy remains feasible despite challenging atmospheric conditions.

Figure~\ref{fig:z7frontier} illustrated the quality of ground-based spectra for the three most distant known quasars. While sufficient to detect broad Mg\,{\sc ii} emission and estimate black hole masses, these spectra suffered from limited wavelength coverage, moderate signal-to-noise, and the inability to access rest-frame optical diagnostics. The H$\beta$ line---better calibrated through local reverberation mapping---remained inaccessible, shifted beyond 4\,$\mu$m where thermal background emission dominates ground-based observations.

\begin{figure}[htbp]
\centering
\includegraphics[width=0.95\textwidth]{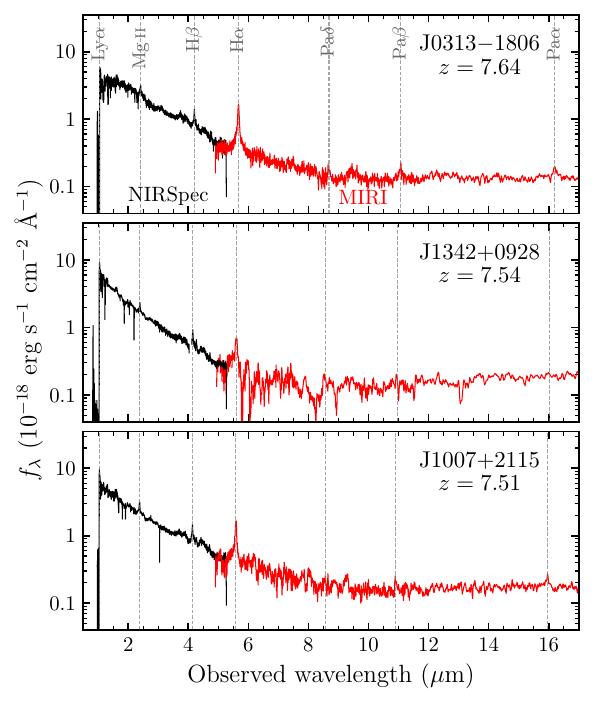}
\caption{The JWST view of the three most distant quasars.  JWST NIRSpec (black) and MIRI (red) spectra of the  $z\sim7.5$ quasars, with key emission lines labelled. The rest-frame optical emission lines that enable direct comparison with lower-redshift calibrations are now accessible. Compare with Figure~\ref{fig:z7frontier}. Reduced spectra courtesy of S.~Bosman, T.~Kist, J.~Hennawi, and J.~Wolf (see also \citealt{bosman2026, kist2026,wolf2026}).} 
\label{fig:jwst-spectra-comparison}
\end{figure}

\runinhead{JWST: Direct Access to Balmer and Paschen Lines.}
JWST's NIRSpec instrument has transformed this situation. Figure~\ref{fig:jwst_capabilities} showed the wavelength coverage available for a $z=7.5$ quasar template. 
Figure~\ref{fig:jwst-spectra-comparison} presents real JWST spectra of the same three $z\sim7.5$ quasars from Fig.~\ref{fig:z7frontier}. NIRSpec's 1--5\,$\mu$m range captures not only Mg\,{\sc ii} but also the [O\,{\sc iii}]$\lambda\lambda$4959,5007 doublet and H$\beta$---the canonical AGN diagnostics that anchor black hole mass calibrations at lower redshift. MIRI extends coverage to the mid-infrared, accessing H$\alpha$ and Paschen lines enabling comprehensive characterisation. 

\begin{figure}[htbp]
\centering
\includegraphics[width=0.7\textwidth]{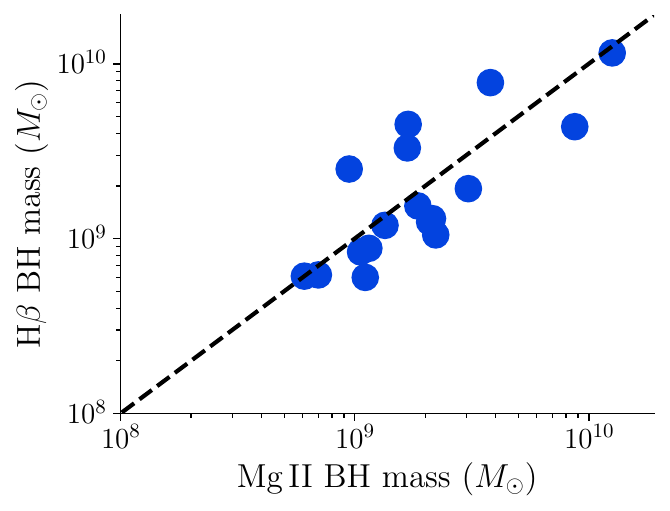}
\caption{Comparison of black hole masses derived from Mg\,{\sc ii} and H$\beta$ for $z>6$ quasars. 
\mgii\ masses are derived from ground-based spectroscopy \citep{fan2023,mazzucchelli2023}; H$\beta$ masses from JWST \citep{Yang2023_aspire,marshall2023,yue2024,loiacono2024}. 
The dashed line shows the one-to-one relation; the scatter is consistent with the intrinsic uncertainty of single-epoch virial estimators.
}
\label{fig:mgii-hbeta-comparison}
\end{figure}

\runinhead{Validating Pre-JWST Mass Estimates.}  An expectation taken directly from section 4.4.1 of \cite{fan2023} was ``\textit{H$\beta$ BH masses are expected to be one of the first results from the first observations of high-redshift quasars with JWST.}'' This is indeed what happened, and now several of the highest redshift quasars count with H$\beta$-based black hole masses. 
The key result from these comparisons is shown in Fig.~~\ref{fig:mgii-hbeta-comparison}: black hole masses derived from Mg\,{\sc ii} agree remarkably well with those derived from H$\beta$. The scatter around the one-to-one relation is consistent with the intrinsic uncertainties in single-epoch mass estimates ($\sim$0.4\,dex), with no evidence for systematic offsets that would indicate catastrophic calibration failures at high redshift.

This validation has profound implications. The existence of $>10^9$\,M$_\odot$ black holes at $z>7$---initially inferred from Mg\,{\sc ii} measurements alone---is now confirmed through independent H$\beta$ diagnostics. The challenge of forming such massive objects within 700\,Myr of the Big Bang, already severe based on pre-JWST estimates, remains as acute as ever. 

\begin{tips}{Notes on Black Hole Mass Calibrations}
The agreement between Mg\,{\sc ii}-based and H$\beta$-based masses is reassuring but should not obscure the underlying uncertainties discussed in Section~\ref{subsec:single_epoch}. The intrinsic scatter in single-epoch relations ($\sim$0.2--0.5\,dex) represents a fundamental limitation for ensemble studies, while individual mass estimates can carry significantly larger uncertainties due to the cumulative effect of each calibration step. What JWST confirms is that the \textit{relative} calibration between Mg\,{\sc ii} and H$\beta$ appears robust, validating statistical conclusions drawn from pre-JWST samples---but not reducing drastically the uncertainty on any individual object.
\end{tips}


\subsection{Kiloparsec Scales: Host Galaxies and Companions}
\label{subsec:kpc}

Having confirmed that the black holes themselves are as massive as pre-JWST estimates suggested, we now examine the galaxies in which they reside.

\subsubsection{The Scientific Motivation}

Quasars are not isolated phenomena but transient phases in the evolution of massive galaxies (Section~\ref{banados-sec:intro}). Understanding the connection between supermassive black holes and their host galaxies at high redshift addresses fundamental questions: Did the black holes form first, with galaxies assembling around them? Did massive galaxies form first, providing the environment for rapid black hole growth? Or did both grow together in a process of co-evolution?

At lower redshifts, tight correlations between black hole mass and host galaxy properties---the $M_{\rm BH}$--$\sigma$ and $M_{\rm BH}$--$M_\ast$ relations---suggest intimate co-evolution. But the origin and redshift evolution of these correlations remain debated \citep[e.g.,][]{habouzit2022}. High-redshift quasars offer the opportunity to test whether these relationships were already in place in the first billion years, or whether early black holes grew independently of (or faster than) their hosts.

\subsubsection{Pre-JWST Knowledge: The Submillimetre View}
\label{subsec:submm}

\begin{figure}[htbp]
\centering
\includegraphics[width=0.85\textwidth]{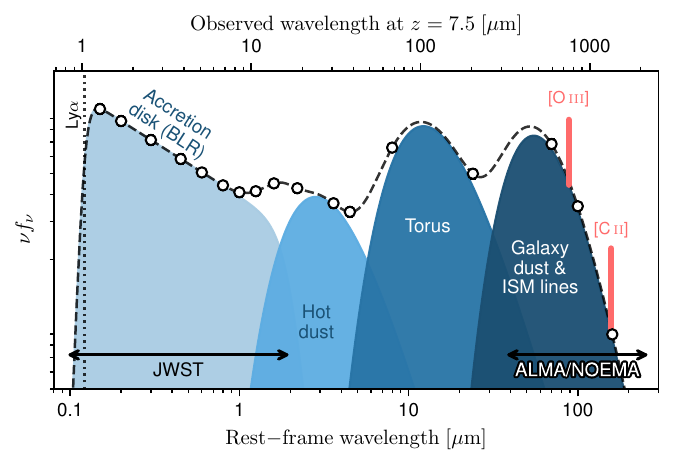}
\caption{Simplified spectral energy distribution of a $z=7.5$ quasar, showing only the dominant components at each wavelength. The central engine---accretion disc, hot dust, and torus---dominates at UV--MIR wavelengths, while host galaxy cold dust dominates in the FIR. Ly$\alpha$ marks the wavelength blueward of which the IGM absorbs nearly all emission (the Gunn--Peterson trough). The [O\,{\sc iii}]\,88\,$\mu$m and [C\,{\sc ii}]\,158\,$\mu$m ISM lines are marked for reference. The top axis shows observed wavelengths: JWST probes the central engine, while ALMA/NOEMA probe the cold dust and ISM. Figure inspired by \citet{leipski2014}.}
\label{fig:quasar-sed-simple}
\end{figure}

Before JWST, our knowledge of quasar host galaxies at $z>6$ came almost exclusively from (sub)millimetre observations \citep[e.g.,][for a detailed recent review see \citealt{decarli2025}]{wang2013,willott2015,banados2015b,izumi2018,decarli2018,2020ApJ...904..130V}. The quasar's central engine dominates at UV through mid-infrared wavelengths, making direct detection of stellar light essentially impossible from the ground. Only at rest-frame far-infrared wavelengths does the host galaxy---specifically its dust continuum and ISM emission lines (e.g., \oiii~88$\mu$m and \cii~158$\mu$m)---become accessible (see Fig.~\ref{fig:quasar-sed-simple}). At $z>6$, these rest-frame FIR wavelengths shift into the (sub)millimetre regime (observed $\sim$0.3--3\,mm), precisely where interferometers like ALMA and NOEMA operate with exceptional sensitivity and angular resolution.

\begin{question}{Historical Perspective: [C\,{\sc ii}] Detections at $z>6$}
When was the first [C\,{\sc ii}] detection on a $z>6$ quasar? The answer is 2005 \citep{maiolino2005}, requiring 12.4\,hours with the IRAM 30m telescope to detect J1148+5251 at $z=6.42$. And what was the previous redshift record? Before this, [C\,{\sc ii}] had only been detected in local galaxies ($z<0.1$). The second $z>6$ [C\,{\sc ii}] detection came seven years later in 2012 \citep{venemans2012}, requiring 6.3\,hours with IRAM/PdBI. What once demanded heroic observational campaigns is now almost routine with ALMA and NOEMA's sensitivity, with more than 100 host galaxy detections achieved to date \citep[e.g.,][]{decarli2018,izumi2019,2020ApJ...904..130V,khusanova2022,banados2024,bouwens2026}.
\end{question}

\begin{figure}[htbp]
\centering
\includegraphics[width=0.7\textwidth]{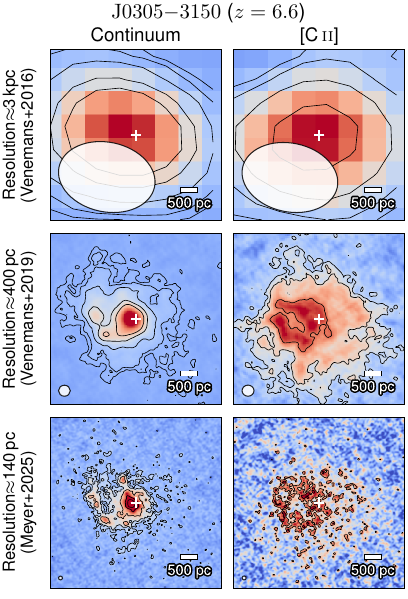}
\caption{ALMA resolves quasar host galaxies at cosmic dawn. The $z=6.6$ quasar J0305$-$3150 observed at three angular resolutions, demonstrating ALMA's capabilities. \textit{Top}: $\sim$3\,kpc resolution. \textit{Middle}: $\sim$400\,pc resolution. \textit{Bottom}: $\sim$140\,pc resolution. The left column shows dust continuum emission; the right column shows [C\,{\sc ii}]\,158\,$\mu$m integrated intensity. The quasar position is marked by white crosses. The synthesized beam is plotted in the bottom left corner of each panel. Increasingly complex morphology emerges at higher resolution. 
Adapted 
from \citet{venemans2016}, \citet{venemans2019}, and \citet{meyer2025}.}
\label{fig:alma-resolution}
\end{figure}

These (sub)millimetre observations have revealed that $z>6$ quasar hosts are massive, gas-rich systems with dynamical masses of $10^{10}$--$10^{11}$\,M$_\odot$ and star formation rates of hundreds of solar masses per year. ALMA's exquisite angular resolution has pushed to $\sim$100\,pc scales, revealing complex morphologies (Fig.~\ref{fig:alma-resolution}; see also \citealt{walter2022,walter2025}). 

However, submillimetre observations trace the cold ISM---gas and dust---rather than the stellar populations that dominate the baryonic mass in present-day galaxies. To directly constrain stellar masses, ages, and morphologies required detecting rest-frame optical/near-infrared emission from stars, which meant waiting for JWST.

\subsubsection{JWST Breakthrough: Seeing the Stars}

\begin{figure}[htbp]
\centering
\includegraphics[width=0.95\textwidth]{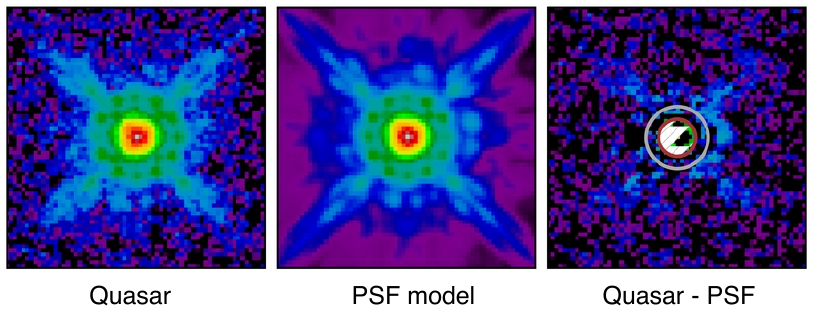}
\caption{\textbf{Pre-JWST attempts to detect stellar light in $z>6$ quasar hosts.} \textit{Left}: HST WFC3/IR image of a $z\sim6$ quasar. \textit{Centre}: Best-fit PSF model. \textit{Right}: Residual after PSF subtraction, showing no significant detection of extended emission. Even with HST's infrared capability and careful PSF modelling, the stellar light from host galaxies remained orders of magnitude below the quasar glare. Figure from  \cite{mechtley2012}. 
}
\label{fig:hst-attempt}
\end{figure}

Detecting stellar light in quasar host galaxies has been a long-standing challenge. Even with HST's infrared capability, the quasar point spread function (PSF) overwhelms any extended emission from the underlying galaxy. Figure~\ref{fig:hst-attempt} illustrates pre-JWST attempts: after careful PSF subtraction, no significant residual emission was detected, placing only upper limits on the host galaxy contribution \citep[e.g.,][]{mechtley2012,marshall2020}.

\begin{figure}[htbp]
\centering
\includegraphics[width=0.95\textwidth]{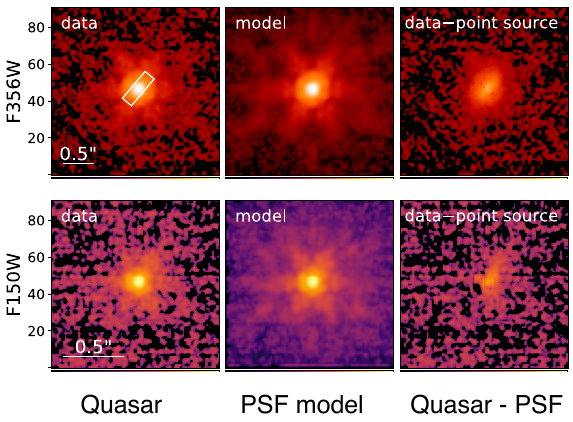}
\caption{\textbf{First detection of stellar light from a $z>6$ quasar host galaxy.} JWST NIRCam imaging of a $z\sim6$ quasar (\textit{left}), PSF model constructed from nearby stars (\textit{centre}), and residual emission after PSF subtraction revealing the host galaxy (\textit{right}). Unlike the unsuccessful HST attempts (Figure~\ref{fig:hst-attempt}), JWST's infrared sensitivity and stable PSF enable routine detection of host galaxy emission. Figure from \citet{ding2023}.}
\label{fig:jwst-host-detection}
\end{figure}

JWST has changed this situation dramatically. Figure~\ref{fig:jwst-host-detection} 
shows one of the first JWST results on $z>6$ quasars of any kind: the detection of 
stellar light from a quasar host galaxy \citep{ding2023}. This paper was posted to 
the arXiv about two weeks after the Cycle~1 observations were taken, showing how 
excited the authors must have been when they finally saw the stars! The combination 
of JWST's larger aperture, superior infrared sensitivity, and remarkably stable PSF 
is making host galaxy detections increasingly common 
\citep[e.g.,][]{yue2024,stone2024,ding2025}.

These detections open a new window on the black hole--host galaxy connection at 
cosmic dawn. Initial results suggest a diversity of host properties, with some 
quasars residing in compact, massive hosts while others show more extended or 
disturbed morphologies indicative of recent interactions (similar diversity has been hinted at by ALMA \cii\ morphologies and kinematics, e.g., \citealt{banados2019a,neeleman2021}). However, current 
detections are typically limited to one or two NIRCam bands per quasar, providing 
only coarse constraints on the host spectral energy distribution. Expanding the 
wavelength coverage will be essential for robust stellar mass and age estimates. 

These measurements also constrain the relationship between black hole mass and host 
galaxy mass at high redshifts. ALMA-based dynamical mass estimates had already 
suggested that $z>6$ quasars are overmassive relative to local scaling relations, 
with $M_\mathrm{BH}/M_\mathrm{dyn}$ ratios roughly an order of magnitude above 
the local $M_\mathrm{BH}$--$M_\mathrm{bulge}$ relation 
\citep{neeleman2021}. JWST stellar mass estimates are now providing 
complementary evidence: initial results from host galaxy detections suggest 
similarly elevated $M_\mathrm{BH}/M_\star$ ratios 
\citep[e.g.,][]{yue2024,stone2024}. Whether this reflects genuine early 
overmassiveness, implying that black holes grew faster than their hosts in the 
first billion years, or is partly driven by selection effects and systematic 
uncertainties in mass estimates, remains actively debated 
\citep[see the detailed discussion in Volonteri's lecture;][]{volonteri2025,li2025}. 
Regardless, the tension is notable: these are among the best-characterised systems 
at $z>6$, with signal-to-noise ratios far exceeding those of the fainter 
JWST-discovered AGN for which similar claims have been made, lending additional 
weight to the overmassiveness finding.

Combining the stellar populations revealed 
by JWST with the cold gas reservoirs mapped by ALMA 
(Section~\ref{subsec:submm}) would provide the first complete picture of the 
baryonic content in these extreme systems. 

\subsubsection{NIRSpec IFU Revelations}

ALMA observations have provided spatially resolved views of quasar host 
galaxies (Sect.~\ref{subsec:submm}) and their companions at submillimetre wavelengths, revealing close 
companion galaxies \citep[e.g.,][]{decarli2017,neeleman2019,pensabene2021} and evidence for outflows 
in the cold gas \citep{bischetti2024}. JWST's NIRSpec IFU now offers a 
complementary view at rest-frame optical wavelengths, probing the ionised gas 
and stellar continuum on similar kiloparsec scales.

By obtaining a spectrum at every spatial position across a $3''\times3''$ field 
of view, the NIRSpec IFU reveals the spatially resolved properties of quasar 
environments in unprecedented detail. The majority of NIRSpec IFU studies so far 
have focused on individual sources or small samples, revealing mergers, outflows, 
and nearby companions in the ionised gas phase 
\citep[e.g.,][]{marshall2023,marshall2025A&A...702A..50M,marshall2025A&A...702A.174M,decarli2024}. 

\begin{figure}[htbp]
\centering
\includegraphics[width=0.95\textwidth]{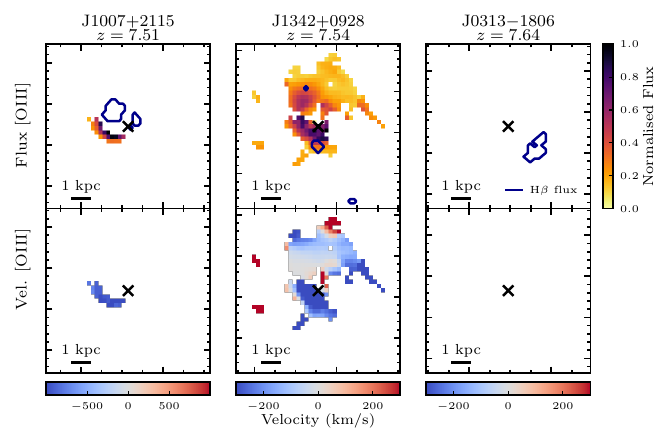}
\caption{[O\,{\sc iii}]$\lambda$5007 flux (\textit{top}) and velocity (\textit{bottom}) maps for the three $z\sim7.5$ quasars with JWST NIRSpec IFU data. Blue contours: H$\beta$ emission; black crosses: quasar positions. Despite similar luminosities and black hole masses, these sources show striking differences: (i)~J1007+2115 exhibits a highly blueshifted ($-870$\,km\,s$^{-1}$) outflow extending $\sim$2\,kpc \citep{liu2024}; (ii)~J1342+0928 shows a $\sim$7\,kpc extended halo \citep{trefoloni2026}; (iii)~J0313$-$1806 shows H$\beta$ but undetected [O\,{\sc iii}] despite 10\,h integration \citep{wolf2026}. Reduced data courtesy of J.~Wolf.  
}
\label{fig:nirspec-ifu-diversity}
\end{figure}

Figure~\ref{fig:nirspec-ifu-diversity} presents [O\,{\sc iii}]$\lambda$5007 flux 
and velocity maps for the three $z\sim7.5$ quasars---J0313$-$1806, J1342+0928, and J1007+2115---protagonists of these lectures 
(see Figs.~\ref{fig:z7frontier} and \ref{fig:jwst-spectra-comparison}). Despite 
their similar luminosities and black hole masses, these sources show striking 
differences in their close environments:

\begin{itemize}
\item \textbf{J1007+2115} exhibits a highly blueshifted ($-870$\,km\,s$^{-1}$) outflow extending $\sim$2\,kpc from the quasar position, with mass outflow rates estimated at 50--300\,M$_\odot$\,yr$^{-1}$ \citep{liu2024}.
\item \textbf{J1342+0928} (Pisco) shows a $\sim$7\,kpc extended [O\,{\sc iii}] halo with relatively quiescent kinematics \citep{trefoloni2026}.
\item \textbf{J0313$-$1806} shows H$\beta$ emission but undetected [O\,{\sc iii}] despite a 10-hour integration, suggesting unusual ionisation conditions or significant dust obscuration \citep{wolf2026}.
\end{itemize}

This diversity is somewhat unexpected. If all luminous $z>7$ quasars formed through similar pathways, one might expect more uniform environments. Instead, the IFU observations suggest that quasar activity can occur in a variety of evolutionary states---from actively merging systems with powerful outflows to more settled configurations with extended gas reservoirs. Understanding what drives this diversity will require larger samples with uniform observations, a key goal for ongoing JWST programmes.


\subsection{Megaparsec Scales: Quasar Environments}
\label{subsec:mpc}

Beyond the immediate surroundings of quasars and their hosts lies the question of 
their large-scale cosmic environment. In hierarchical structure formation, the most 
massive dark matter haloes at any epoch are expected to host the most extreme 
objects. Luminous quasars, requiring both massive black holes and abundant gas 
supplies, should therefore trace the highest peaks in the primordial density field. 
Simulations predict that $z>6$ quasars inhabit haloes with masses 
$M_{\rm halo} \gtrsim 10^{12}$\,M$_\odot$, surrounded by overdensities of lower-mass 
galaxies \citep{angulo2012,costa2014,costa2024}. But do the observations support this picture?

\subsubsection{Pre-JWST Searches}

Testing this prediction proved surprisingly difficult. Two decades of searching for 
galaxy overdensities around high-redshift quasars yielded mixed results---only two robust 
spectroscopic confirmations of Mpc-scale overdensities existed before JWST 
(see Section 5.5 of \citealt{fan2023}, and their Figure 11).  

The challenges were both observational and astrophysical. On the observational side, 
spectroscopic confirmation of faint galaxies at $z>6$ required heroic efforts with 
ground-based telescopes, severely limiting sample sizes. On the astrophysical side, 
quasar environments may be genuinely diverse: some theoretical models predict that 
AGN feedback can suppress star formation in the quasar's immediate vicinity, 
creating ``voids'' of galaxies rather than overdensities at small scales \citep[][]{habouzit2019,chen2020ApJ...893..165C}.

\begin{figure}[htbp]
\centering
\includegraphics[width=0.85\textwidth]{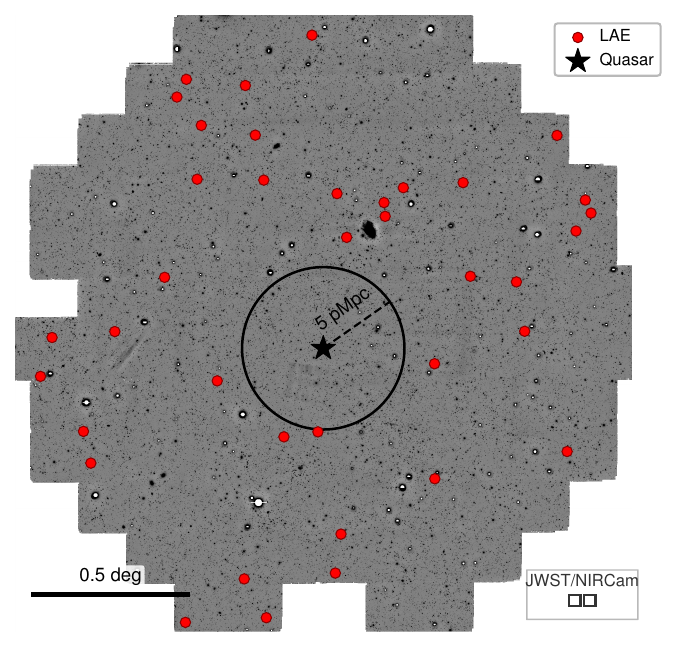}
\caption{Distribution of Lyman-$\alpha$ emitters around the $z=6.9$ quasar J2348$-$3054 from DECam narrowband imaging ($\sim$3\,deg$^2$). A notable deficit of LAEs within 5\,pMpc contrasts with an overdensity at larger radii. The JWST/NIRCam field of view (lower right) is shown for reference.  Adapted from \citet{lambert2024}.}
\label{fig:lae-search}
\end{figure}

A recent study illustrates this complexity. \citet{lambert2024} searched for 
Lyman-$\alpha$ emitters (LAEs) around the $z=6.9$ quasar J2348$-$3054, covering 
approximately 3\,deg$^2$ with DECam narrowband imaging (Fig.~\ref{fig:lae-search}). 
The result was puzzling: a deficit of LAEs within $\sim$5\,Mpc of the quasar, but a 
significant overdensity at larger distances. Whether this reflects AGN feedback 
suppressing local galaxy formation or simply the large cosmic variance expected at 
high redshift remains unclear. This study also demonstrates that ground-based 
wide-field surveys remain valuable even alongside JWST: the overdensity revealed by 
\citet{lambert2024} lies on scales beyond JWST's field of view 
(Fig.~\ref{fig:lae-search}, lower right).

\subsubsection{JWST Slitless Spectroscopy: The Game Changer}

JWST's NIRCam grism mode overcomes the key pre-JWST limitation: the need for 
targeted, time-intensive spectroscopy of individual galaxy candidates. Instead, 
slitless spectroscopy captures spectra of every source in the field of view 
simultaneously, enabling blind identification of [O\,{\sc iii}] emitters without 
pre-selecting targets. Two major programmes---ASPIRE 
\citep[25 quasar fields at $6.5<z<6.8$;][]{wang2023ApJ...951L...4W,champagne2025} and EIGER \citep[4 quasar fields at $6.0<z<6.5$;][]{kashino2023,eilers2024}---are 
now systematically mapping the environments of $z>6$ quasars using this technique.

\begin{figure}[htbp]
\centering
\includegraphics[width=0.8\textwidth]{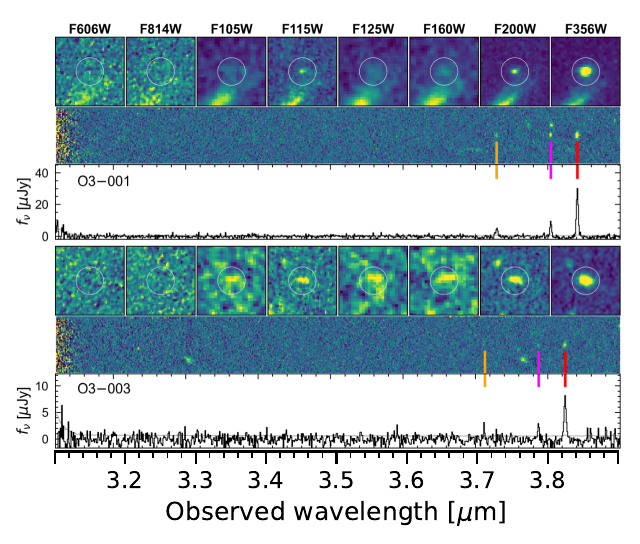}
\caption{{Example [O\,{\sc iii}] emitter detection from JWST slitless spectroscopy.} \textit{Top}: Multi-band image cutouts ($2''\times2''$) from HST (F606W through F160W) and JWST (F115W, F200W, F356W). The galaxy is undetected in optical HST bands as expected for $z>5.7$. \textit{Middle}: 2D coadded grism spectrum showing clear [O\,{\sc iii}] doublet emission (and sometimes H$\beta$). \textit{Bottom}: Optimally extracted 1D spectrum with H$\beta$, [O\,{\sc iii}]$\lambda$4959, and [O\,{\sc iii}]$\lambda$5007 marked. Adapted from the ASPIRE survey \citep{wang2023ApJ...951L...4W}.}
\label{fig:slitless-detection}
\end{figure}

Figure~\ref{fig:slitless-detection} illustrates the method. Galaxies that would 
require hours of targeted spectroscopy are identified directly from the grism data, 
with redshifts determined from the [O\,{\sc iii}] doublet. The 
efficiency is remarkable: a single NIRCam pointing yields spectroscopic redshifts 
for dozens of emission-line galaxies across a redshift range of $5 \lesssim z 
\lesssim 7$.

\begin{figure}[htbp]
\centering
\includegraphics[width=0.95\textwidth]{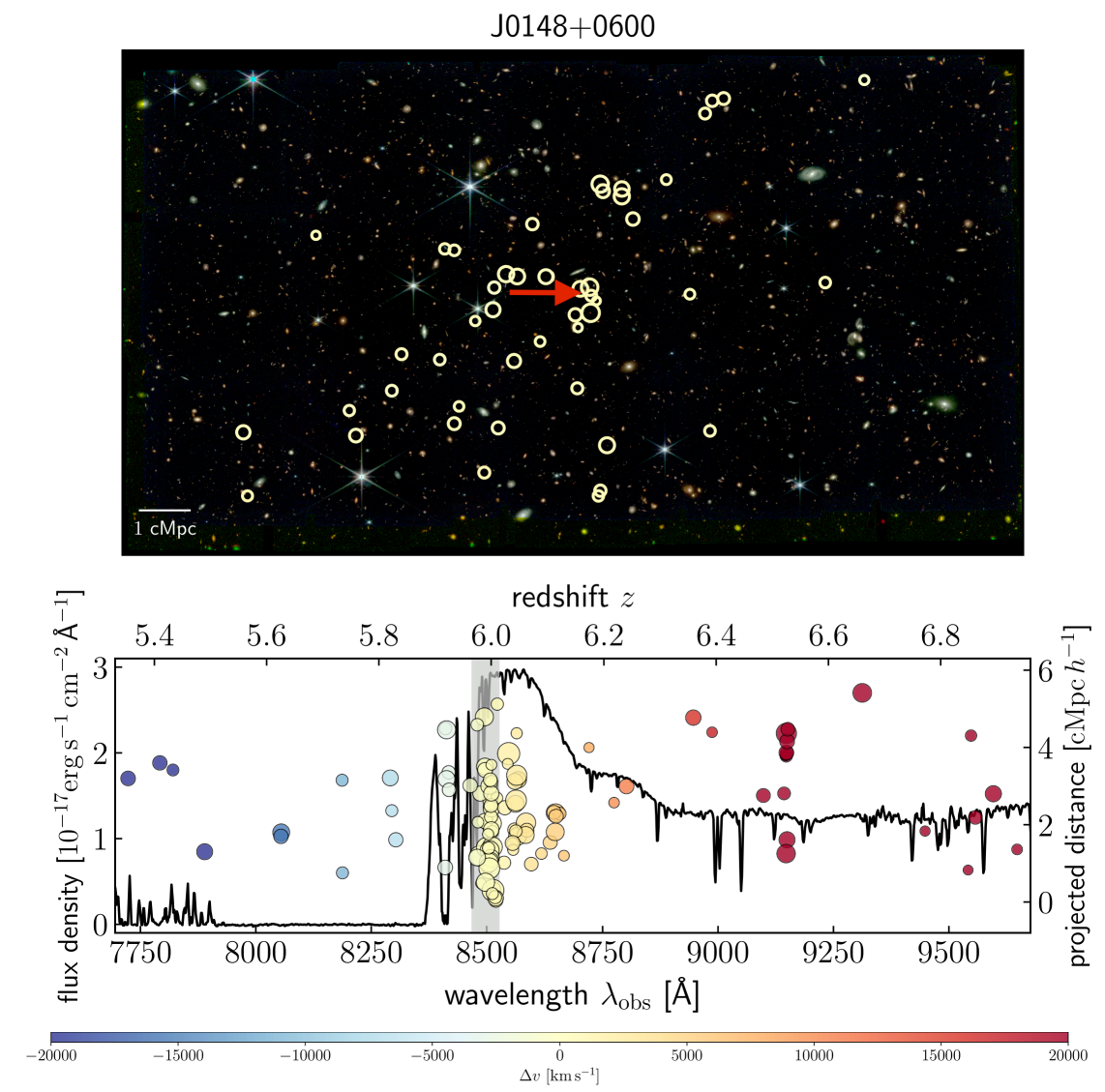}
\caption{JWST reveals quasar environments through slitless spectroscopy. \textit{Top}: NIRCam RGB image (F115W, F200W, F356W) of the $z=6.0$ quasar J0148+0600 field, with the quasar marked by a red arrow. Circles highlight [O\,{\sc iii}] emitters within $|\Delta v| < 1000$\,km\,s$^{-1}$ of the quasar redshift. \textit{Bottom}: Ground-based quasar spectrum with [O\,{\sc iii}] emitters overplotted; circle size scales with $\log L_{\rm [O\,III]}$ and colour indicates velocity offset relative to the quasar (colour bar at bottom). The field shows a substantial overdensity of galaxies near the quasar redshift (grey shaded region), as well as foreground and background structure, demonstrating JWST's ability to map three-dimensional large-scale structure around early quasars. From the EIGER survey \citep{eilers2024}.}
\label{fig:aspire-eiger-environments}
\end{figure}

Figure~\ref{fig:aspire-eiger-environments} shows an example of the resulting 
science: the field around the $z=6.0$ quasar J0148+0600, displaying the largest overdensity 
of [O\,{\sc iii}] emitters at the quasar redshift. Through clustering analysis of 
such fields, the EIGER programme has derived the first constraint on quasar host 
dark matter halo masses at $z>6$, finding 
$\log(M_{\rm halo}/{\rm M}_\odot) > 12.4$ \citep{eilers2024}---consistent with 
quasars inhabiting some of the most massive haloes at this epoch, as predicted by 
simulations.

However, the field-to-field variance is enormous. In my lectures, I discussed 
preliminary results from the ASPIRE survey hinting at this diversity; as I was 
finishing these lecture notes, the full analysis appeared 
\citep{wang2026arXiv260204979W}, and I could not resist including it here. The 
halo mass lower limits derived from ASPIRE are consistent with the EIGER results, 
but reveal that cosmic variance is substantial \citep{wang2026arXiv260204979W,huang2026}.

\begin{figure}[htbp]
\centering
\includegraphics[width=0.99\textwidth]{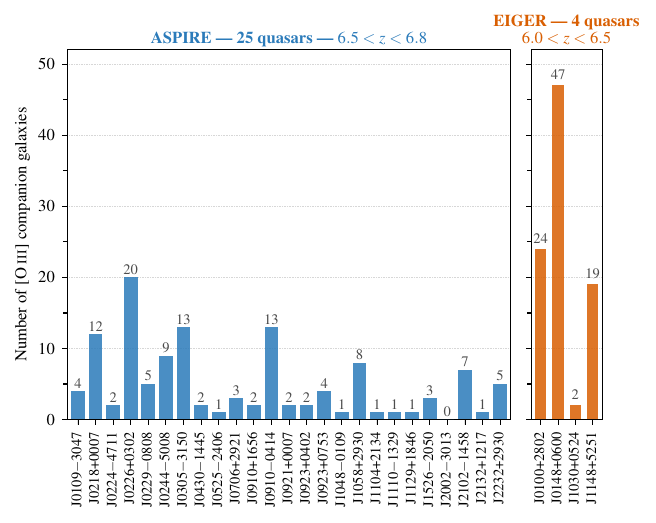}
\caption{Large field-to-field variance in quasar companion galaxies. The number of 
[O\,{\sc iii}] emitters within $|\Delta v| < 1000$\,km\,s$^{-1}$ of each quasar 
ranges from zero to 20 in ASPIRE (25 fields) and from 2 to 47 in EIGER (4 fields). 
The higher EIGER counts partly reflect its $2\times2$ NIRCam mosaic design, 
covering $\sim$60\% more area than single ASPIRE pointings, and greater depth. 
While quasars trace overdense environments on average, the scatter is enormous. 
Data from \citet{wang2026arXiv260204979W} and \citet{eilers2024}.}
\label{fig:aspire-eiger-variance}
\end{figure}

Figure~\ref{fig:aspire-eiger-variance} quantifies this scatter: the number of 
[O\,{\sc iii}] companion galaxies ranges from zero to 20 across the 25 ASPIRE 
fields, and from 2 to 47 in the four EIGER fields. While quasars trace overdense 
environments on average, some fields show no excess companions at all---a result 
reminiscent of the LAE deficit found by \citet{lambert2024} on larger scales 
(Fig.~\ref{fig:lae-search}). Whether these ``empty'' fields reflect AGN feedback 
suppressing nearby galaxy formation, unfavourable sight-line geometry, or simply 
the stochastic nature of early structure formation remains an open question.

\subsubsection{Little Red Dots}

The clustering measurements just described also enable a comparison with the emerging population of ``Little Red Dots'' (LRDs). LRDs are addressed in detail in the lectures by Richard Ellis and Marta Volonteri \citep{ellis2025,volonteri2025}. Here 
I note only their relevance to the environmental studies discussed above.

\begin{figure}[htbp]
\centering
\includegraphics[width=0.95\textwidth]{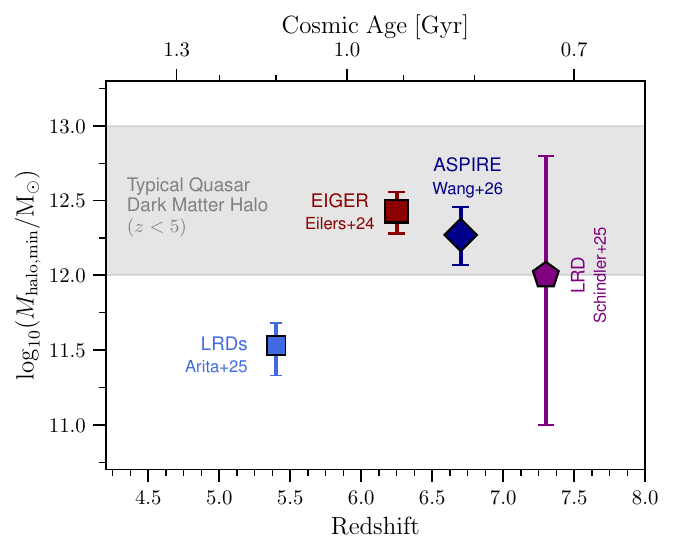}
\caption{Minimum dark matter halo mass as a function of redshift for quasars and 
Little Red Dots (LRDs) at $z > 5$. The grey shaded region indicates the typical 
halo mass range for UV-luminous quasars at $z < 5$. At $z > 6$, quasars from EIGER 
\citep{eilers2024} and ASPIRE \citep{wang2026arXiv260204979W} occupy similarly 
massive haloes. The LRD population at $z \sim 5.4$ \citep{arita2025} resides in 
haloes approximately one dex less massive, suggesting a distinct population. 
However, one LRD at $z = 7.3$ \citep{schindler2025} has a halo mass consistent with 
quasars, possibly indicating an obscured quasar rather than a typical LRD.}
\label{fig:dm_halo_clustering}
\end{figure}

A natural question is whether luminous quasars and LRDs represent the same 
population observed in different evolutionary states, or distinct phenomena 
entirely. Comparing their dark matter halo masses offers one test. 
Figure~\ref{fig:dm_halo_clustering} compiles current constraints: the EIGER and 
ASPIRE surveys find that $z>6$ quasars inhabit haloes with 
$\log(M_{\rm halo,min}/{\rm M}_\odot) \gtrsim 12$--12.5 
\citep{eilers2024,wang2026arXiv260204979W}, consistent with the extrapolation from 
lower-redshift quasar clustering measurements (grey band). In contrast, 
\citet{arita2025} find that LRDs at $z \sim 5.4$ reside in haloes approximately one 
order of magnitude less massive, suggesting they may represent a distinct 
population.

The picture is not entirely clear, however. \citet{schindler2025} study the 
environment of a single LRD at $z = 7.3$ and find a halo mass consistent with 
UV-luminous quasars, raising the possibility that this source is an obscured quasar 
rather than a typical LRD. This highlights a key uncertainty: LRDs likely comprise 
multiple populations with different physical origins, and environmental studies may prove essential for 
disentangling them.

\subsection{Looking Forward and Concluding Remarks}
\label{subsec:future}

This section has examined JWST's impact on our understanding of early supermassive 
black holes across three orders of magnitude in physical scale. At parsec scales, 
JWST spectroscopy has largely \textit{confirmed} the picture developed through two 
decades of ground-based work: the Mg\,{\sc ii}-based black hole masses agree 
remarkably well with JWST H$\beta$ measurements, and the early Universe does indeed 
host billion-solar-mass black holes by $z \sim 7$. At kiloparsec scales, JWST has 
delivered genuine \textit{breakthroughs}, including the first detections of stellar light in 
quasar host galaxies, and striking diversity in the environments revealed by 
NIRSpec IFU observations. At megaparsec scales, slitless spectroscopy programmes 
are mapping large-scale structure around high-redshift quasars with unprecedented 
completeness, confirming that these objects inhabit massive dark matter haloes 
while revealing enormous field-to-field variance.

These results set the stage for the coming decade. But perhaps the most  important 
change is not any single measurement. It is how JWST is bringing previously separate communities together.

\subsubsection{The Converging Communities}

The high-redshift universe looks remarkably different depending on which community 
meeting you attend. At quasar conferences, participants present diagrams like 
Figure~\ref{fig:quasar-census-update}---showing the census of the most distant 
quasars, with the current highest redshift sources at $z\sim 7.5$. At galaxy 
conferences, the view is quite different, with JWST now discovering galaxies up to 
$z\sim 14$ \citep{carniani2024,naidu2025arXiv250511263N}. Richard Ellis's lectures 
in this volume show the dramatic improvement in the $z>6$ spectroscopic redshift 
dataset, with JWST contributing over 1300 confirmed sources where pre-JWST efforts 
had secured barely 60 (see Fig.~23 in \citealt{ellis2025}).

\begin{figure}[htbp]
\centering
\includegraphics[width=1\textwidth]{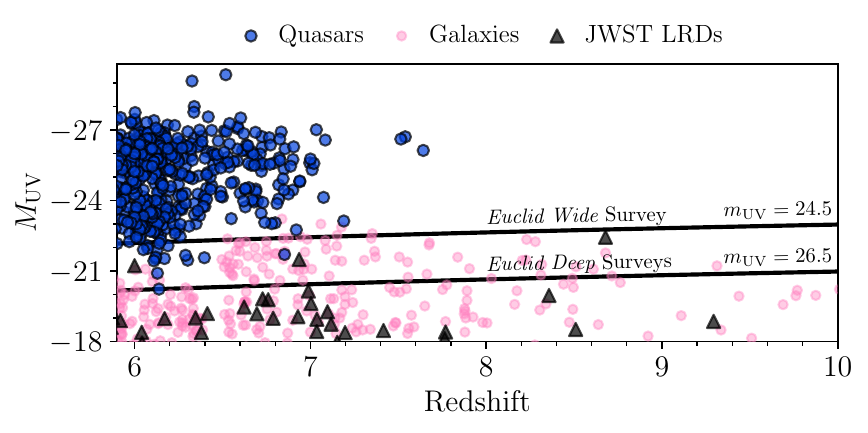}
\caption{UV magnitude versus redshift for spectroscopically confirmed quasars, 
galaxies, and JWST Little Red Dots (LRDs) in the first Gyr of the Universe. 
(Quasars: see Sect.~\ref{subsec:quasar-landscape} and Fig.~\ref{fig:quasar-census-update}; 
Galaxies: \citealt{bouwens2022,roberts-borsani2024,roberts-borsani2025a};
LRDs: \citealt{harikane2023,larson2023,maiolino2024,degraaff2026}.) At 
\textit{Euclid}'s depths (horizontal lines for Wide and Deep surveys), UV-faint 
quasars overlap with luminous galaxies and LRDs. This convergence is motivating 
much more frequent dialogue between the galaxy and quasar communities.}
\label{fig:zuv-combined}
\end{figure}

Figure~\ref{fig:zuv-combined} brings both quasar and galaxy populations into the 
same $M_{\rm UV}$ vs.\ redshift parameter space, along with the emerging population 
of ``Little Red Dots'' (LRDs) discussed in the lectures by Richard Ellis and Marta 
Volonteri \citep{ellis2025,volonteri2025}. We are entering an era where these 
populations overlap in luminosity---and where the boundary between ``quasar'' and 
``galaxy'' becomes increasingly blurred.

This convergence will only accelerate. The \textit{Euclid} mission, launched in 
July 2023, is delivering all-sky near-infrared imaging that bridges the gap between 
luminous quasars and fainter AGN populations. At the time of my lectures in January 
2025, \textit{Euclid} had just released its first data, and the results were as 
good as we had hoped (and I know for a fact that Fig.~\ref{fig:zuv-combined} will need an update very soon). The potential for discovering quasars at even higher redshifts, perhaps $z\sim10$, is tantalising. The \textit{Nancy Grace Roman Space 
Telescope} and Vera C.\ Rubin Observatory will further expand this parameter space, 
while JWST continues to probe the faintest sources in unprecedented detail. By early 2030s, ESO's Extremely Large Telescope (ELT) will reach similar depths to JWST but at much higher spatial and spectral resolution. 
The ELT will expand direct black hole mass measurements into the epoch of reionization, enable detailed morphological studies of distant galaxies, and trace early chemical enrichment through absorption systems towards high-redshift quasars.

\subsubsection{Final Reflections}

As one of my mentors once reminded me, astronomy is fundamentally about people. The progress described in this chapter reflects the work of hundreds of researchers who built the ground-based surveys, developed the mass measurement techniques, pushed ALMA and NOEMA to their limits, planned and built JWST and \textit{Euclid} over decades, and are now analysing the flood of data these facilities produce.

The collaboration between historically separate communities---quasar astronomers 
and galaxy astronomers, optical observers and radio observers---is producing 
insights that neither group could achieve alone. As Figure~\ref{fig:zuv-combined} 
illustrates, the artificial boundary between ``quasar'' and ``galaxy'' populations 
is dissolving at cosmic dawn. The most important discoveries ahead will come from these communities working together. 

\begin{important}{Take-Home Messages}
\begin{enumerate}
\item \textbf{Black hole masses are validated:} JWST H$\beta$ measurements confirm 
Mg\,{\sc ii}-based masses from ground-based spectroscopy. The existence of 
$>10^9$\,M$_\odot$ black holes at $z>7$ is now secure, making the formation puzzle 
even more pressing.

\item \textbf{Host galaxy stellar light is detectable:} After decades of 
non-detections, JWST reveals the stellar populations of quasar hosts for the first 
time. Early results suggest diverse stellar masses and morphologies.

\item \textbf{Close environments show striking diversity:} NIRSpec IFU observations 
reveal that quasars of similar luminosity can inhabit very different 
kiloparsec-scale environments, from quiescent systems to active mergers with 
powerful outflows.

\item \textbf{Large-scale environments are overdense on average:} Slitless 
spectroscopy surveys confirm that $z>6$ quasars inhabit massive dark matter haloes 
($\log M_{\rm halo}/{\rm M}_\odot \gtrsim 12$), though with enormous field-to-field 
variance.

\item \textbf{The communities are converging:} As JWST reveals lower-luminosity AGN 
and \textit{Euclid} probes the faint quasar regime and bright galaxy regime, the boundary between ``quasar'' 
and ``galaxy'' populations is dissolving, enabling a unified view of massive black 
holes at cosmic dawn. \textit{This very school, bringing together lecturers and students from both communities to discuss galaxies and black holes in the first billion years as a single subject, is a reflection of that convergence.}
\end{enumerate}
\end{important}

\begin{acknowledgement}
I thank the organisers of the 54th Saas-Fee Advanced Course---Romain Meyer, Michaela Hirschmann, Pascal Oesch---for the invitation to lecture at 
this winter school. Having my family there to share this beautiful corner of the world made it all the more memorable. 

I was fortunate to sahre the school with my fellow lecturers, whose work has haped parts of my own research, as hopefully noted in parts of this book. While I had crossed paths 
with Richard Ellis and Marta Volonteri before, this was the first time I met Rachel 
Somerville in person. It was a pleasure to 
finally connect.

The students were outstanding. I had prepared interactive lectures with 
frequent questions, and I worried that I might be met with silence. Those 
worries were unfounded: there was no shortage of raised hands, and the 
discussions were often the highlight of each session. In the year since the school, I have run into several of you at conferences and meetings. It is always good to see what you 
are working on, and I look forward to crossing paths again. 

I thank the colleagues who generously shared data and results that made several figures in this book (and lectures) possible. 
Laura Mart\'inez-Ram\'irez created Fig.~\ref{fig:agnsed}. Raphael Hviding provided the spectra for Figs.~\ref{fig:unification} and \ref{fig:neon}. 
Sarah Bosman, Timo Kist, Joe Hennawi, and Julien Wolf provided reduced JWST data to create Figs.~\ref{fig:jwst-spectra-comparison} and \ref{fig:nirspec-ifu-diversity}. Bram Venemans, Romain Meyer, and Fabian Walter provided the ALMA data to create Fig.~\ref{fig:alma-resolution}. Trystan Lambert provided the DECAM data and codes to create Fig.~\ref{fig:lae-search}. Feige Wang kindly allowed me to present the ASPIRE clustering results during the school well before publication.  Rebecca Bowler provided a list of confirmed galaxies I had originally missed in Fig.~\ref{fig:zuv-combined}.

I also thank the mentors and colleagues who helped shape these lectures from the start. The high-redshift group at MPIA sat through my first attempt at the opening lecture and provided invaluable feedback. The Saas-Fee students surely benefited from their input. Members of both the MPIA high-redshift group and the Heidelberg AGN community also provided comments on these written notes. For their time and insights, I thank Silvia Belladitta, Joyce Glass, Raphael Hviding, Laura 
Mart\'inez-Ram\'irez, Jörg-Uwe Pott, Francisco Pozo Nu\~nez, and Zhang-Liang Xie.
\end{acknowledgement}

\bibliographystyle{spbasic}
\bibliography{banados_bibliography}

\end{document}